\documentclass[fleqn,usenatbib]{stile/mnras}

\usepackage{txfonts}
\usepackage[T1]{fontenc}
\usepackage{ae,aecompl}

\usepackage{graphicx}

\usepackage{amsmath}
\usepackage{amsfonts}
\usepackage{amssymb}

\usepackage{xcolor}

\usepackage{breakcites}
\usepackage{hyperref}
\usepackage{footnote}

\usepackage{comment}

\usepackage{xfrac}        % use packages statements

\def\nat{Nature}

\def\mnras{MNRAS}
\def\apj{ApJ}
\def\apjl{ApJL}
\def\apjs{ApJS}
\def\aap{A\&A}

\def\araa{ARA\&A}
\def\aj{AJ}
\def\qjras{QJRAS}
\def\physrep{Phys. Rep.}
\def\pasj{Pub. Astron. Soc. Japan}

\newcommand{\quotes}[1]{``#1''}

\newcommand{\angstrom}{\textup{\AA}}

\def\gtsima{$\; \buildrel > \over \sim \;$}
\def\ltsima{$\; \buildrel < \over \sim \;$}
\def\gsim{\lower.5ex\hbox{\gtsima}}
\def\lsim{\lower.5ex\hbox{\ltsima}}
\def\gtsima{$\; \buildrel > \over \sim \;$} 
\def\ltsima{$\; \buildrel < \over \sim \;$} \def\gsim{\lower.5ex\hbox{\gtsima}} 
\def\lsim{\lower.5ex\hbox{\ltsima}} 
\def\simgt{\lower.5ex\hbox{\gtsima}} 
\def\simlt{\lower.5ex\hbox{\ltsima}}

\def\be{\begin{equation}}
\def\ee{\end{equation}}

\def\NII{\hbox{[N~$\scriptstyle\rm II $]}}

\def\Hb{\hbox{H~$ \beta $~}}
\def\Ha{\hbox{H~$ \alpha $~}}

\def\OII{\hbox{[O~$\scriptstyle\rm II $]}}
\def\CII{\hbox{[C~$\scriptstyle\rm II $]}}

\def\NIII{\hbox{[N~$\scriptstyle\rm III $]}}
\def\OIII{\hbox{[O~$\scriptstyle\rm III $]}}

\graphicspath{{plots/}}                 % folder(s) with plots

\begin{document}

\title[Cosmic $T_{\rm d}$ evolution out to $z \sim 7$]{\centering The REBELS ALMA Survey:\\ \centering cosmic dust temperature evolution out to $z \sim 7$}

\author[Sommovigo et al.]{L. Sommovigo$^{1}$\thanks{\href{mailto:laura.sommovigo@sns.it}{laura.sommovigo@sns.it}},
A. Ferrara$^{1}$, A. Pallottini$^{1}$, P. Dayal$^{2}$,
R.J. Bouwens$^{3}$, R. Smit$^{4}$, E. da Cunha$^{5,6b}$, \newauthor 
I. De Looze$^{7,8}$, R. A. A. Bowler$^{9}$, 
J. Hodge$^{3}$, H. Inami$^{10}$, P. Oesch$^{11,12}$, R. Endsley$^{13}$, V. Gonzalez$^{14,15}$,\newauthor  
S. Schouws$^{3}$, D. Stark$^{13}$, M. Stefanon$^{3}$, M. Aravena$^{16}$, L. Graziani$^{17,18}$,
D. Riechers$^{19}$, \newauthor
R. Schneider$^{17,20}$, P. van der Werf$^{3}$, H. Algera$^{10}$, L. Barrufet$^{11}$, Y. Fudamoto$^{11,21,22}$,\newauthor 
A. P. S. Hygate$^{3}$, I. Labb{\'e}$^{23}$, Y. Li$^{24,25}$, T. Nanayakkara$^{23}$, M. Topping$^{13}$\\
% list of institutions
$^{1}$ Scuola Normale Superiore, Piazza dei Cavalieri 7, I-56126 Pisa, Italy\\
$^{2}${Kapteyn Astronomical Institute, University of Groningen, 9700 AV Groningen, The Netherlands}\\
$^{3}${Leiden Observatory, Leiden University, NL-2300 RA
  Leiden, Netherlands}\\ 
$^{4}${Astrophysics Research
  Institute, Liverpool John Moores University, 146 Brownlow Hill,
  Liverpool L3 5RF, United Kingdom}\\ 
$^{5}${International Centre for
  Radio Astronomy Research, University of Western Australia, 35
  Stirling Hwy, Crawley,26WA 6009, Australia}\\ 
$^{6}$ARC Centre of Excellence for All Sky Astrophysics in 3 Dimensions (ASTRO 3D), Australia\\
$^{7}${Sterrenkundig Observatorium, Ghent University,
  Krijgslaan 281 - S9, 9000 Gent, Belgium} \\ 
$^{8}${Dept. of
  Physics \& Astronomy, University College London, Gower Street,
  London WC1E 6BT, United Kingdom}\\    
$^{9}${Astrophysics, The
  Denys Wilkinson Building, University of Oxford, Keble Road, Oxford,
  OX1 3RH, United Kingdom}\\   
$^{10}${Hiroshima Astrophysical Science Center, Hiroshima
  University, 1-3-1 Kagamiyama, Higashi-Hiroshima, Hiroshima 739-8526,
  Japan}\\   
$^{11}${Observatoire de Gen{\`e}ve, 1290 Versoix,
  Switzerland}\\  
$^{12}${Cosmic Dawn Center (DAWN), Niels Bohr
  Institute, University of Copenhagen, Jagtvej 128, K\o benhavn N,
  DK-2200, Denmark}\\  
$^{13}${Steward Observatory,
  University of Arizona, 933 N Cherry Ave, Tucson, AZ 85721, United
  States}\\  
$^{14}${Departmento de Astronomia, Universidad de
  Chile, Casilla 36-D, Santiago 7591245, Chile}\\ 
$^{15}${Centro de Astrofisica y Tecnologias Afines (CATA),
  Camino del Observatorio 1515, Las Condes, Santiago, 7591245, Chile}\\ 
$^{16}${Nucleo de Astronomia, Facultad
  de Ingenieria y Ciencias, Universidad Diego Portales, Av. Ejercito
  441, Santiago, Chile}\\  
$^{17}${Dipartimento di
  Fisica, Sapienza, Universita di Roma, Piazzale Aldo Moro 5, I-00185
  Roma, Italy}\\  
$^{18}${INAF/Osservatorio Astrofisico di
  Arcetri, Largo E. Femi 5, I-50125 Firenze, Italy}\\ 
$^{19}${I. Physikalisches Institut, Universit\"at zu K\"oln, Z\"ulpicher 
Strasse 77, D-50937 K\"oln, Germany} \\ 
$^{20}${INAF/Osservatorio
  Astronomico di Roma, via Frascati 33, 00078 Monte Porzio Catone,
  Roma, Italy}\\  
$^{21}${Research Institute for Science and Engineering,
  Waseda University, 3-4-1 Okubo, Shinjuku, Tokyo 169-8555, Japan}\\ 
$^{22}${National Astronomical Observatory of Japan, 2-21-1,
  Osawa, Mitaka, Tokyo, Japan}\\  
$^{23}${Centre for Astrophysics \&
  Supercomputing, Swinburne University of Technology, PO Box 218,
  Hawthorn, VIC 3112, Australia}\\  
$^{24}${Department of Astronomy \& Astrophysics, The
  Pennsylvania State University, 525 Davey Lab, University Park, PA
  16802, USA}\\  
$^{25}${Institute for Gravitation and the
  Cosmos, The Pennsylvania State University, University Park, PA
  16802, USA}\\  
}

\label{firstpage}
\pagerange{\pageref{firstpage}--\pageref{lastpage}}

\maketitle

\begin{abstract}
ALMA observations have revealed the presence of dust in the first generations of galaxies in the Universe. However, the dust temperature $T_{\rm d}$ remains mostly unconstrained due to the few available FIR continuum data at redshift $z>5$. This introduces large uncertainties in several properties of high-$z$ galaxies, namely their dust masses, infrared luminosities, and obscured fraction of star formation.
Using a new method based on simultaneous \CII\ 158$\mu$m line and underlying dust continuum measurements, we derive $T_{\rm d}$ in the continuum and \CII\ detected $z\approx 7$ galaxies in the ALMA Large Project REBELS sample.  We find $39\ \mathrm{K} < T_{\rm d} < 58\ \mathrm{K}$, and dust masses in the narrow range $M_{\rm d} = (0.9-3.6)\times 10^7 M_\odot$. 
These results allow us to extend for the first time the reported $T_{\rm d}(z)$ relation into the Epoch of Reionization. We produce a \textit{new physical model} that explains the increasing $T_{\rm d}(z)$ trend with the decrease of gas depletion time, $t_{\rm dep}=M_{\rm g}/{\rm SFR}$, induced by the higher cosmological accretion rate at early times; this hypothesis yields $T_{\rm d}
\propto (1+z)^{0.4}$. The model also explains the observed $T_{\rm d}$ scatter at a fixed redshift. We find that dust is warmer in obscured sources, as a larger obscuration results in more efficient dust heating. For UV-transparent (obscured) galaxies, $T_{\rm d}$ only depends on the gas column density (metallicity), $T_{\rm d} \propto N_{\rm H}^{1/6}$ ($T_{\rm d} \propto Z^{-1/6}$). 
REBELS galaxies are on average relatively transparent, with effective gas column densities around $N_{\rm H} \simeq (0.03-1)\times 10^{21} {\rm cm}^{-2}$. We predict that other high-$z$ galaxies (e.g. MACS0416-Y1, A2744-YD4), with estimated $T_{\rm d} \gg 60$ K, are significantly obscured, low-metallicity systems. In fact $T_{\rm d}$ is higher in metal-poor systems due to their smaller dust content, which for fixed $L_{\rm IR}$ results in warmer temperatures.
\end{abstract}


\section{Introduction}
The rest-frame Ultraviolet (UV) emission from galaxies in the Epoch of Reionization (EoR) has been extensively studied thanks to the Hubble Space Telescope (HST) and ground-based telescopes 
(for UV luminosity functions see: \citealt{2012Brad,Oesch_2013,2013McLure,2015Bowler,Atek15,2015Bouwens,2017Liverm}, for single detections see: \citealt{2010Bouw,2011Bouw,Ellis_2012,2014Brad,Oesch_2016}).

Recently, also the Far-Infrared (FIR) emission from such early sources has become observable thanks to the advent of high sensitivity millimetre interferometers such as the Atacama Large Millimeter Array (for surveys see: \citealt{capak15,Carilli_2016,Bouwens16,barisic2017dust,bowler2018obscured,bethermin_alpine,Schaerer_alpine}, for comprehensive reviews see: \citealt{2013carilli,2016Dunlop}). 
The combination of data from these different instruments has allowed us to glimpse the interstellar medium (ISM) of early galaxies (see e.g. \citealt{10.1093/mnras/stt702,Stark16,Dayal18,2020HodgeCuhna}) for the first time.

The FIR emission from galaxies includes FIR lines, arising from atomic and molecular species in the ISM, and dust continuum emission.  
One of the brightest (and thus most commonly observed) FIR lines is the fine-structure transition of singly ionized carbon \CII\ $158\mathrm{\mu m}$, which traces mainly the neutral atomic gas in the ISM \citep{1991ApJ...373..423S,RevModPhys.71.173,Wolfire_2003}. The dust continuum is the thermal radiation emitted by dust grains heated by the UV and optical light coming from young stars \citep[see e.g.][]{draine1989interstellar,meurer1999dust,calzetti2000dust,weingartner2001dust,Draine03}. 

Dust grains span a wide range of physical temperatures deepening on their own physical properties and the radiation field heating them \citep[see e.g.][]{Draine03}. Nevertheless, under the approximation of thermal equilibrium, the FIR spectral energy distribution (SED) can be well approximated by an isothermal grey-body function \citep{1983QJRAS..24..267H}. Where available, observations at rest-frame Mid-Infrared (MIR) wavelengths, short-wards of the FIR emission peak, have shown significant deviations from a grey-body \citep{Dunne01,Blain03,Kovacs_2010,2012casey,Dale_2012,Galam12,Kirkpatrick_2012,2008Cuhna,Cunha_2015,casey2018,ReuterSPT}. Unfortunately, MIR wavelengths are inaccessible at $z>5$ with current instruments. Given the limited availability of observations sampling the full dust emission regime, the most common and physically-sound approach is to assume isothermal dust emitting as a grey-body. The key dust properties constrained through SED fitting are the dust temperature $T_{\rm d}$ and mass $M_{\rm d}$\footnote{An additional parameter is the dust emissivity, $\beta_{\rm d}$; see  Section \ref{method}.}, which are degenerate quantities.

Reliably determining high-$z$ galaxies $T_{\rm d}$ and $M_{\rm d}$ holds the key to many problems related to early galaxy formation and evolution. For  example, it might shed light on the heating produced by obscured star formation, and on the nature and processes governing the dust formation and content in the early universe

Interestingly, in the last few years several works spanning the redshift range $0 \leq z \leq 6$ have suggested the presence of a direct correlation between $T_{\rm d}$ and redshift \citep[see e.g.][]{2012Magdis,2013Magnelli,2015Bet,Schreiber18,dusttemp2020,Bouwens20,ReuterSPT}. 
Combining the few available dust temperature estimates for individual galaxies at $z>5$ with lower-$z$ stacking results, different works came to discrepant conclusions. While \cite{Bouwens20} have confirmed the reported linear $T_{\rm d}(z)$ increase, \cite{dusttemp2020} suggested instead a flattening of the relation at higher redshift. A physical and quantitative interpretation of either trend is still lacking. Previous works suggested that warm dust temperatures at high-$z$ might result from a more compact dust geometry in high-$z$ galaxies (w.r.t. local galaxies), with dust being mostly located in the vicinity of young stars, and thus being more efficiently heated \citep{Liang19,sommovigo20}. Other works have suggested that the increasing $T_{\rm d}(z)$ trend can be qualitatively ascribed to the growing specific star formation rate at high-$z$ \citep{2014A&A...561A..86M,Ma16,2019MNRAS.487.1844M,2021arXiv210412788S}, together with the lower dust content of early galaxies \citep[see e.g.][]{2017MNRAS.466..105A,behrens18,2021arXiv210412788S,pallottini:2022}. A lower dust content would in fact result in warmer dust temperatures for a given FIR luminosity $L_{\rm FIR}$.

Firmer conclusions on the $T_{\rm d}$ evolution at early times have been hindered by two factors. First, the number of galaxies at $z>5$ for which multiple FIR continuum observations are available is very limited \citep{big3drag,behrens18,2020Harik,Bakx20,dusttemp2020}. 
In fact, most ALMA sources, when detected in dust continuum, have only a single (or very few) data point(s) at FIR wavelengths \citep[e.g.][]{2016knud,Bouwens16,2016Pav,barisic2017dust,bowler2018obscured,2019Pav,big3drag,Tamura19,bethermin_alpine}. Hence a value for $T_{\rm d}$ is often assumed \textit{a priori} in the fitting procedure (or associated to very large experimental uncertainties).
The second limiting factor is the very large scatter of measured $T_{\rm d}$ values, ranging from $T_{\rm d} \simeq 25\ \mathrm{K}$ \citep[][]{2020Harik} up to $T_{\rm d}>80\ \mathrm{K}$ in the narrow redshift range $z= 6.2-8.3$ \citep[][]{Bakx20}. 

The lack of (or poor) knowledge of $T_{\rm d}$ at high-$z$ results in very large uncertainties on $M_{\rm d}$ as well as derived galaxy properties, such as infrared luminosity ($L_{\rm IR}$) and obscured star formation rate \citep[SFR; see e.g.][]{sommovigo20}. In the future, the problem can be mitigated thanks to further ALMA observations in multiple higher-frequency bands (Bands 7,8,9). In fact, for galaxies at $z \simgt 5$ such observations sample the SED closer to the emission peak in the FIR \citep[see][]{bakx:2021}.

To overcome the current FIR observational limitations at $z\ge 5$, in \cite{TdCIImetodo} we developed a new method aimed at simultaneously constraining $T_{\rm d}$ and $M_{\rm d}$ with a single band measurement. The main idea is to combine the widely observed fine-structure \CII\ 158$\mu$m line with the underlying dust continuum emission at the same frequency ($1900\ \mathrm{GHz}$). 

Our method can improve the reliability of the interpretation of \CII\ and continuum observations from millimetre interferometers. This is particularly relevant in the context of recent ALMA large programs targeting \CII\ emitters at $z\simgt 5$, such as the very recent \textit{\quotes{Reionization Era Bright Emission Line Survey}} (REBELS; PI: Bouwens, \citealt{bouwens:2021}).
REBELS studied $40$ of the brightest known galaxies at $z>6.5$ identified over a $7\ \mathrm{deg}^2$ area of the sky, systematically scanning for bright ISM-cooling lines, \CII $158\mathrm{\mu m}$ and \OIII $88\mathrm{\mu m}$, and dust-continuum emission \citep[for further details see also][]{Fudamoto_2021}.

REBELS galaxies are UV-selected sources at redshift $z= 6.5-7.7$ ($-21.3 \simlt M_{\rm UV} \simlt -22.5$); they are luminous galaxies with stellar masses\footnote{derived from rest-frame UV SED fitting by \cite{Stefanon:2022} using BEAGLE \citep{10.1093/mnras/stw1756}. The authors adopt a constant Star Formation History (SFH), $0.2\ \mathrm{Z_{\odot}}$ metallicity, a \cite{calzetti2000dust} dust extinction law, and a \cite{2003ApJ...586L.133C} $0.1-300\ \mathrm{M_{\odot}}$ initial mass function. Note that the correction on the given $M_{\star}$ values, required to be consistent with a Salpeter $1-100\ \mathrm{M_{\odot}}$ IMF (used in the rest of the paper), is well within their uncertainties and does not affect significantly our results. We caution that using a non-parametric prescription for the SFH might result in $M_{\star}$ values up to a factor $\sim \times 3$ larger (for a detailed discussion on this point see \citealt{topping:2022}). Implications on the results presented here are discussed in the text.} in the range $10^9 \simlt M_{\rm \star}/M_{\odot} \simlt 10^{10}$, and relatively high SFRs around $20 \simlt \mathrm{SFR}/\mathrm{M_{\odot}yr^{-1}} \simlt 200$ \citep{bouwens:2021}. 

Here we apply our method \citep{TdCIImetodo} to the $13$ REBELS targets detected both in continuum ($3 \sigma$) and in \CII\ ($5 \sigma$) at $z=6.5-8.5$ (for the sources properties see Table \ref{tabREB}). They represent the first statistical sample of continuum detections at such early epochs, featuring a four-fold increase of the size of the previously available galaxy sample.

Constraining the dust temperatures and masses of REBELS galaxies is crucial for a better understanding of SF obscuration and dust production in these early galaxies. Moreover, our analysis on the REBELS sample can investigate for the first time the reported cosmic dust temperature evolution \citep{2012Magdis,2013Magnelli,2015Bet,Schreiber18,dusttemp2020,Bouwens20} into the Epoch of Reionization. In particular, the eventual flattening in the $T_{\rm d}(z)$ evolution would be more robustly identified at the high ($z \approx 7$) redshift of the REBELS sample.

This paper is organised as follows. 
In  Section \ref{method} we summarize the \citet[][]{TdCIImetodo} method to compute $T_{\rm d}$, which is then applied to REBELS galaxies in  Section \ref{apllreb}.  Section \ref{tdredev} is devoted to the analysis of the reported $T_{\rm d}$-redshift relation in the light of the derived REBELS results. There we introduce a new physical model to explain the observed increasing $T_{\rm d}-z$ trend. In  Section \ref{summary} a summary and discussion of the results is given.

Throughout the paper, we assume a $\Lambda$CDM model with the following cosmological parameters: $\Omega_{\rm M} = 0.3075$, $\Omega_{\Lambda} = 1- \Omega_{\rm M}$, $\Omega_{\rm B} = 0.0486$,  $h=0.6774$, and $\sigma_8=0.81$. $\Omega_{\rm M}$, $\Omega_{\Lambda}$, $\Omega_{\rm B}$ are the total matter, vacuum, and baryonic densities, in units of the critical density; $h$ is the Hubble constant in units of $100\, {\rm km s}^{-1}$, and $\sigma_8$ is the late-time fluctuation amplitude parameter \citep{Planck16}.

\begin{table*}
    \caption{\textbf{On the left}: Measured REBELS galaxies properties, respectively: redshift $z$, stellar mass $\log M_{\star}$, \CII\ luminosity $L_{\rm CII}$, and $1900\ \mathrm{GHz}$ continuum flux $F_{\rm 1900}$ (not CMB-corrected). For the data analysis we refer to the dedicated papers by \citealt{inami:2022} ($F_{\rm 1900}$), \citealt{sander:2022} ($z,L_{\rm CII}$, see also \citealt{bouwens:2021}), and \citealt{Stefanon:2022} ($M_{\star}$). \textbf{On the right}: Predicted REBELS galaxies properties, respectively: \CII-to-total gas conversion factor $\alpha_{\rm CII}$ (eq. \ref{alfahz}), dust temperature $T_{\rm d}$ and mass $M_{\rm d}$, IR luminosity $\log L_{\rm IR}$, SN dust yield $y_d$ and obscured SFR, $\mathrm{SFR}_{\rm IR} [M_\odot {\rm yr}^{-1}] = 10^{-10}\ \mathrm{L_{\rm IR}} [L_\odot]$ \citep{kennicutt1998}. Please note that the galaxy names in the text are abbreviated as REBxx for conciseness.}
  \begin{center}
    \begin{tabular}{l|c|c|c|c|c|c|c|c|c|c}
        \hline\hline
        \multicolumn{4}{c}{\textit{Measured}}& \multicolumn{1}{c}{ID\#} & \multicolumn{6}{c}{\textit{Predicted}}\\
        \hline
$z$&  $\log M_{\rm \star}$& $L_{\rm CII}$ & $F_{\rm 1900}$ &   & $\alpha_{\rm CII}$ & $T_{\rm d}$& $\log M_{\rm d}$ &$\log L_{\rm IR}$ & $y_{\rm d}$ & $\mathrm{SFR}_{\rm IR}$\\\hline
  & $[M_\odot]$          & $[10^8\, L_\odot]$ & $[\mu$Jy]   &   & & [K]          & $[M_\odot]$       & $[L_\odot ]$ & $\mathrm{[M_{\odot}]}$ & $\mathrm{[M_{\odot} yr^{-1}]}$\\\hline

6.496 & $9.2^{+0.9}_{-1.0}$ & $ 6.9 \pm 0.4 $ & $ 67 \pm 13 $   & REB05 & $ 5 ^{+ 4 }_{- 2 }$ & $ 43 ^{+ 15 }_{- 9 }$ & $ 7.14 ^{+ 0.31 }_{- 0.3 }$ & $ 11.37 ^{+ 0.5 }_{- 0.32 }$ & $ 0.51 ^{+ 0.52 }_{- 0.26 }$ & $ 23 ^{+ 51 }_{- 12 }$\\

6.749 & $9.0^{+0.6}_{-0.7}$ & $ 7.4 \pm 0.9 $ & $ 101 \pm 20 $   & REB08 & $ 5 ^{+ 3 }_{- 2 }$ & $ 50 ^{+ 17 }_{- 11 }$ & $ 7.17 ^{+ 0.25 }_{- 0.25 }$ & $ 11.83 ^{+ 0.5 }_{- 0.36 }$ & $ 0.74 ^{+ 0.58 }_{- 0.33 }$ & $ 68 ^{+ 147 }_{- 39 }$\\

7.346 & $8.9^{+0.9}_{-0.7}$ & $ 10.1 \pm 1.9 $ & $ 87 \pm 24 $   & REB12 & $ 4 ^{+ 2 }_{- 1 }$ & $ 49 ^{+ 16 }_{- 9 }$ & $ 7.19 ^{+ 0.22 }_{- 0.26 }$ & $ 11.79 ^{+ 0.49 }_{- 0.33 }$ & $ 0.93 ^{+ 0.6 }_{- 0.41 }$ & $ 62 ^{+ 130 }_{- 33 }$\\

7.084 & $8.7_{-0.7}^{+0.8}$ & $ 3.7 \pm 0.5 $ & $ 60 \pm 15 $   & REB14 & $ 6 ^{+ 3 }_{- 2 }$ & $ 52 ^{+ 15 }_{- 10 }$ & $ 6.95 ^{+ 0.22 }_{- 0.23 }$ & $ 11.67 ^{+ 0.44 }_{- 0.34 }$ & $ 0.87 ^{+ 0.59 }_{- 0.36 }$ & $ 47 ^{+ 84 }_{- 25 }$\\

7.675 & $9.5_{-0.7}^{+0.6}$ & $ 10.8 \pm 0.7 $ & $ 53 \pm 10 $   & REB18 & $ 4 ^{+ 5 }_{- 2 }$ & $ 39 ^{+ 12 }_{- 7 }$ & $ 7.30 ^{+ 0.34 }_{- 0.32 }$ & $ 11.29 ^{+ 0.4 }_{- 0.21 }$ & $ 0.34 ^{+ 0.4 }_{- 0.18 }$ & $ 19 ^{+ 30 }_{- 7 }$\\

7.370 & $8.8_{-0.7}^{+0.7}$ & $ 8.7 \pm 1.7 $ & $ 71 \pm 20 $   & REB19 & $ 3 ^{+ 2 }_{- 1 }$ & $ 50 ^{+ 16 }_{- 9 }$ & $ 7.09 ^{+ 0.19 }_{- 0.24 }$ & $ 11.74 ^{+ 0.48 }_{- 0.32 }$ & $ 1.05 ^{+ 0.57 }_{- 0.45 }$ & $ 55 ^{+ 110 }_{- 28 }$\\

7.307 & $9.9_{-0.2}^{+0.2}$ & $ 15.9 \pm 0.4 $ & $ 260 \pm 22 $   & REB25 & $ 5 ^{+ 5 }_{- 2 }$ & $ 55 ^{+ 15 }_{- 14 }$ & $ 7.55 ^{+ 0.32 }_{- 0.22 }$ & $ 12.45 ^{+ 0.43 }_{- 0.45 }$ & $ 0.24 ^{+ 0.27 }_{- 0.1 }$ & $ 284 ^{+ 480 }_{- 184 }$\\

7.090 & $9.7_{-0.3}^{+0.3}$ & $ 6.1 \pm 0.6 $ & $ 51 \pm 10 $   & REB27 & $ 5 ^{+ 7 }_{- 2 }$ & $ 41 ^{+ 15 }_{- 9 }$ & $ 7.14 ^{+ 0.37 }_{- 0.32 }$ & $ 11.25 ^{+ 0.48 }_{- 0.3 }$ & $ 0.15 ^{+ 0.2 }_{- 0.08 }$ & $ 18 ^{+ 37 }_{- 9 }$\\

6.685 & $9.6_{-0.2}^{+0.2}$  & $ 5.5 \pm 0.9 $ & $ 56 \pm 13 $   & REB29 & $ 6 ^{+ 7 }_{- 2 }$ & $ 42 ^{+ 16 }_{- 10 }$ & $ 7.11 ^{+ 0.37 }_{- 0.32 }$ & $ 11.27 ^{+ 0.53 }_{- 0.35 }$ & $ 0.16 ^{+ 0.22 }_{- 0.09 }$ & $ 19 ^{+ 44 }_{- 10 }$\\

6.729 & $9.6_{-0.4}^{+0.4}$ & $ 7.9 \pm 0.6 $ & $ 60 \pm 17 $   & REB32 & $ 5 ^{+ 6 }_{- 2 }$ & $ 39 ^{+ 15 }_{- 9 }$ & $ 7.21 ^{+ 0.35 }_{- 0.32 }$ & $ 11.22 ^{+ 0.51 }_{- 0.31 }$ & $ 0.24 ^{+ 0.31 }_{- 0.13 }$ & $ 17 ^{+ 37 }_{- 9 }$\\

6.577 & $9.6_{-1.3}^{+0.7}$ & $ 16.8 \pm 1.3 $ & $ 163 \pm 23 $   & REB38 & $ 4 ^{+ 4 }_{- 2 }$ & $ 46 ^{+ 18 }_{- 11 }$ & $ 7.45 ^{+ 0.32 }_{- 0.3 }$ & $ 11.89 ^{+ 0.55 }_{- 0.38 }$ & $ 0.39 ^{+ 0.43 }_{- 0.2 }$ & $ 77 ^{+ 195 }_{- 45 }$\\

6.845 & $8.6_{-0.6}^{+0.6}$ & $ 7.9 \pm 1.4 $ & $ 80 \pm 16 $   & REB39 & $ 3 ^{+ 1 }_{- 1 }$ & $ 58 ^{+ 15 }_{- 9 }$ & $ 6.95 ^{+ 0.13 }_{- 0.19 }$ & $ 11.98 ^{+ 0.4 }_{- 0.29 }$ & $ 1.3 ^{+ 0.46 }_{- 0.45 }$ & $ 95 ^{+ 145 }_{- 46 }$\\

7.365 & $9.5_{-1.0}^{+0.5}$ & $ 4.9 \pm 1.0 $ & $ 48 \pm 13 $   & REB40 & $ 6 ^{+ 7 }_{- 3 }$ & $ 43 ^{+ 17 }_{- 10 }$ & $ 7.07 ^{+ 0.37 }_{- 0.32 }$ & $ 11.34 ^{+ 0.54 }_{- 0.34 }$ & $ 0.21 ^{+ 0.27 }_{- 0.11 }$ & $ 22 ^{+ 54 }_{- 12 }$\\

      \hline\hline
    \end{tabular}
    \label{tabREB}
\end{center}
\end{table*}

\section{Method}\label{method}

In \cite{TdCIImetodo} we proposed a novel method to derive the dust temperature in galaxies, based on the combination of the $1900\ \mathrm{GHz}$ continuum and the super-imposed \CII\ line emission. We summarize the method in the following.

We use the \CII\ luminosity, $L_{\rm CII}$, as a proxy for the total gas mass $M_{\rm g}$, or equivalently the dust mass $M_{\rm d}$, given a dust-to-gas ratio $D$:
\begin{equation}\label{Mdust}
M_{\rm d} = D M_{\rm g} = D\, \alpha_{\rm CII} L_{\rm CII},
\end{equation}
where $\alpha_{\rm CII}$ is the \CII-to-total gas conversion factor. 

We derive an analytic expression for $\alpha_{\rm CII}$ using empirical relations such as the Kennicutt–Schmidt relation \citep[][hereafter, KS]{kennicutt1998}, and the De Looze relation between $L_{\rm CII}-$SFR \citep[][hereafter, DL]{delooze14}. This yields
\begin{equation}\label{alfahz}
    \alpha_{\rm CII}=32.47\ \frac{y^2}{\kappa_{\rm s}^{5/7}}\  \Sigma_{\rm SFR}^{-0.29} \quad \frac{M_\odot}{L_\odot}\,,
\end{equation}
where $\Sigma_{\rm SFR}$ is the SFR surface density, and $\kappa_{\rm s}$ is the \quotes{burstiness parameter} \citep[][]{ferraraCII,2019MNRAS.487.1689P,Vallini20} which quantifies upwards deviations from the KS relation ($\kappa_{\rm s}>1$ for starbursts\footnote{$\kappa_{\rm s}=1$ for normal galaxies, being defined as $\kappa_s = \Sigma_{\rm SFR}/(10^{-12}\ \Sigma_{\rm gas}^{1.4})$ \citep{Heiderman_2010}}). The factor $y=r_{\rm CII}/r_{\star}>1$ is introduced since there is growing evidence \citep[$1.5 \simlt y \simlt 3$ at $z > 4$, see e.g.][]{carniani:2017oiii,carniani2018clumps,matthee2017alma,Matthee_2019,2019ApJ...887..107F,2019Ryb,2020arXiv200300013F,2020A&A...633A..90G,Carniani20} that at $z>4$ \CII\, emission (size $r_{\rm CII}$) is more extended than the UV size ($r_{\star}$).

We assume the dust-to-gas ratio $D$ to scale linearly with metallicity down to $Z \simlt 0.1\ Z_{\odot}$ \citep{james2002, 2007ApJ...657..810D,galliano2008,Leroy_2011,2014RR}:
\begin{equation}\label{DtoZ}
    D=D_{\odot}\ \left(\frac{Z}{Z_{\odot}}\right)\,,
\end{equation}
where $D_{\odot}=1/162$ is the Galactic value \citep{2014RR}. This simple linear $D-Z$ relation is confirmed to hold in simulated galaxies at $z\sim 7$ for stellar masses in the range $10^{9} < M_{\star}/M_{\odot} < 10^{11}$ (\citealt{dayal:2022}, see also \citealt{Ma16,Torrey19}).

Armed with an expression for $\alpha_{\rm CII}$, $D$, and thus $M_{\rm d}$ (eq. \ref{Mdust}), we can constrain $T_{\rm d}$ using the continuum flux at $1900\ \mathrm{GHz}$. We consider Milky Way-like dust\footnote{Milky Way-like dust seems the best suited to reproduce high-$z$ sources properties, see \citep{bowler2018obscured,2021arXiv210512133S,ferrara:2022} for a detailed discussion.}, for which standard values for the dust opacity\footnote{This power-law approximation for the dust opacity is valid at wavelengths $\lambda >20\ \mathrm{\mu m}$, well within the FIR range.} $\kappa_{\nu} = \kappa_{\star} (\nu/\nu_{\star})^{\beta_{\rm d}}$ are $(\kappa_{\star}, \nu_{\star}, \beta_{\rm d})$ = (10.41 ${\rm cm^2 g^{-1}}$, $1900\, {\rm GHz}$, $2.03$) \citep{weingartner2001dust,Draine03}.

In \cite{TdCIImetodo} we rewrite the equation for the continuum flux emitted by a given dust mass\footnote{assuming an isothermal, optically-thin dust emission model} with (CMB-corrected) equilibrium temperature $T_{\rm d}$, yielding the following explicit expression for $T_{\rm d}$:
\begin{equation}\label{td1pto}
T_{\rm d} = \frac{T_{0}}{\ln(1+f^{-1})}\,,
\end{equation}
where $T_0= h_{\rm P} \nu_0/ k_{\rm B} = 91.86$ K is the temperature corresponding to the \CII\ transition energy at $\nu_0=1900$ GHz, $k_{\rm B}$ and $h_{\rm P}$ are the Boltzmann and Planck constants, respectively. The function $f$ is defined as: 
\begin{equation}\label{f}
f = {\cal B}(T_{\rm CMB}) + A^{-1}\tilde F_{\nu_0},
\end{equation}
where ${\cal B}(T_{\rm CMB})= [\exp(T_0/T_{\rm CMB})-1)]^{-1}$, and $T_{\rm CMB}$ is the CMB temperature at a given redshift. The non-dimensional continuum flux $\tilde F_{\nu_0}$ and the constant\footnote{Note that the value given here is $20\%$ lower w.r.t. the one given in \cite{TdCIImetodo} due to the slightly different dust model} $A$ correspond to:
\begin{equation}\label{defs}
\begin{split}
    &\tilde F_{\nu_0} = 0.98 \times 10^{-16} \left(\frac{F_{\nu_0}}{\rm mJy} \right),\\
    & A =  4.33 \times 10^{-24} \left[\frac{g(z)}{g(6)}\right] \left(\frac{L_{\rm CII}}{L_\odot}\right)\left(\frac{\alpha_{\rm CII}}{M_\odot/L_\odot}\right)  D\,,\\
\end{split}
\end{equation}
where $g(z) = {(1+z)}/{d_L^2}$, and $d_{\rm L}$ is the luminosity distance to redshift $z$. Eq. \ref{td1pto} can be used to compute $T_{\rm d}$ using a single $1900$ GHz observation (which provides both $L_{\rm CII}$ and $F_{ \nu_0}$) modulo an estimate for $D$ (eq. \ref{DtoZ}) and $\alpha_{\rm CII}$ (eq. \ref{alfahz}). 

Writing explicitly the expressions for $D$ and $\alpha_{\rm CII}$ it turns out that $T_{\rm d}$ is ultimately a function of $(\kappa_{\rm s},z,F_{\nu_ 0},Z,\Sigma_{\rm SFR},L_{\rm CII},y) $. All these parameters are generally constrained by observations out to very high redshift with two exceptions. Both $\kappa_{\rm s}$ and the metallicity $Z$ are largely unknown. Hence in the derivation we consider a broad random uniform distribution for both parameters (see  Section \ref{apllreb}). To optimally constrain $T_{\rm d}$, we add the following broad physical conditions: 
\begin{itemize}
    \item[$\square$] $M_{\rm d}$ cannot exceed the maximal dust production per supernova (SN), $M_{\rm d, max} = 0.04\ M_{\star}$\footnote{Note that to set this broad upper limit we do not account for dust destruction.}. This expression for $M_{\rm d, max}$ is obtained by assuming a standard Salpeter $1-100\ \mathrm{M_{\odot}}$ Initial Mass Function \citep[IMF, ][]{10.1046/j.1365-8711.2000.03209.x}, and that all the SN metal yield ($\simeq 2\ \mathrm{M_{\odot}}$ per SN) gets locked into dust grains \citep[see eq. 15 in][]{TdCIImetodo}.
    \item[$\square$] The IR-deduced star formation, $\mathrm{SFR_{\rm IR}} = 10^{-10}\ L_{\rm IR}$ \citep{kennicutt1998}, cannot largely exceed the SFR deduced from \CII\ using the DL relation for starbursts\footnote{Precisely, we allow $\mathrm{SFR}_{\rm IR}$ to deviate at most by $ \simlt 1\ \mathrm{dex}$ from $\mathrm{SFR}/ (\mathrm{M_{\odot} yr^{-1}}) = 10^{-7.06}\ L_{\rm CII}/\mathrm{L_{\odot}}$ \citep{delooze14}, which is comparable to the dispersion around this relation observed at high-$z$ \citep[see e.g.][]{Carniani20}}.
\end{itemize}
We recall that the relation
\begin{equation}\label{LIR}
{L_{\rm IR}}=\left( \frac{M_{\rm d}}{M_{\odot}} \right)\ \left( \frac{T_{\rm d}}{8.5\, \rm K} \right)^{4+\beta_{\rm d}} L_{\odot}.    
\end{equation}
holds for the dust model adopted here \citep{ferrara:2022}. Solutions not satisfying both these two bounds are discarded. These conditions result in a lower (upper) cut on $T_{\rm d}$ corresponding to unphysically large dust masses (FIR luminosities/SFR). This allows us to effectively constrain $T_{\rm d}$ at high-$z$ despite of the lack of information on $(\kappa_{\rm s},Z)$.

This method for the dust temperature derivation has been tested on a sample of $19$ local galaxies and four galaxies at $z \simgt 4$ \citep{TdCIImetodo,bakx:2021}. For all these galaxies multiple data points in the FIR SED are available, allowing us to compare our inferred dust temperatures with robust $T_{\rm d}$ estimates obtained with traditional SED fitting. For the $19$ local galaxies all the parameters $(\kappa_{\rm s},z,F_{\nu_ 0},Z,\Sigma_{\rm SFR},L_{\rm CII},y)$ are constrained by observations. This is also true for $3$ out of the $4$ high-$z$ galaxies considered in the test\footnote{For reference, in the considered local galaxies: $0.05 \simlt Z/Z_{\odot} \simlt 2.75$, $0.1 \simlt \kappa_{\rm s} \simlt 5.9$ and \CII\ emitting regions are as extended as stars ($y\sim 1$). For the $3$ high-$z$ galaxies SPT0418-47, MACS416-Y1 and B14-65666: $0.2 \simlt Z/Z_{\odot} \simlt 0.4$, $9 \simlt \kappa_{\rm s} \simlt 45$ and $1.2 \simlt y \simlt 1.5$.}. For the remaining high-$z$ galaxy, A1689-zD1 at $z=7.13$, both $(\kappa_{\rm s},Z)$ are unknown hence we assume the same broad distributions of values considered here for REBELS galaxies ($Z = 0.3-1\ Z_{\odot}$ and $\kappa_{\rm s} = 1-50$). These assumptions are motivated in detail in Sec. \ref{apllreb}. Encouragingly, we recovered consistent dust temperatures with traditional SED fitting  within $1 \sigma$ spanning the very large redshift range $z = 0-8.31$ (as well as the large temperature range $20\ \mathrm{K} \simlt T_{\rm d} \simlt 100\ \mathrm{K}$). 

We also tested our method on simulations, applying it to the $z \sim 6.7$ galaxy Zinnia (a.k.a. serra05:s46:h0643) from the SERRA simulation suite \citep[][see also \citealt{pallottini:2022}]{2019MNRAS.487.1689P}. Also in this case, we recover $T_{\rm d}$ in agreement with single-temperature grey body SED fitting performed at the frequencies corresponding to ALMA bands 6, 7, and 8. 

In \citep{TdCIImetodo} we have shown that the dust temperature derived from our method matches that obtained from FIR SED fitting. We underline that this single value, which we refer to as $T_{\rm d}$ throughout the paper, does not necessarily correspond to the dust physical temperature, which is instead characterised by a Probability Distribution Function (PDF, see e.g. \citealt{behrens18,sommovigo20}). 

In general, $T_{\rm d}$ does not necessarily provide a statistically sound representation of the PDF. We investgated the relation between the two in Appendix A of \cite{TdCIImetodo} for the simulated galaxy Zinnia. At $z \simgt 5$ such detailed comparison with observations is not currently possible due to the lack of information on the galaxies dust temperature PDF. In the case of Zinnia, we found that $T_{\rm d}$ corresponds to the galaxy mass-weighted dust temperature ($\sim 60\ \mathrm{K}$), which is a factor $\sim 2\times$ colder than the luminosity-weighted dust temperature. In fact, Zinnia shows an excess of emission at MIR wavelengths (which is not traced by $T_{\rm d}$) due to the presence of a scarce, but very luminous, hot dust component. The generalization of such PDF-$T_{\rm d}$ relation is pending on a thorough comparison with a larger number of simulated galaxies, and future instruments observations possibly providing MIR-to-IR coverage for $z \simgt 5$ galaxies (such as the yet to be launched space telescopes Millimetron, see e.g. \citealt{Novikov_2021}, and Origins, see e.g. \citealt{2021ExA...tmp..117W}).

Finally, we remark that the increase by a factor $\sim 3$ in the stellar masses, which might result from using a non parametric SFH \citep{topping:2022}, barely affects our results. In fact, the derived values of $\alpha_{\rm CII}$, $M_{\rm d}$, and $T_{\rm d}$ of individual galaxies remain nearly unaffected, given their uncertainties, even by varying the stellar-mass based upper limits on $M_{\rm d}$ by that amount.

\begin{figure*}
    \centering
    \includegraphics[width=1.0\linewidth]{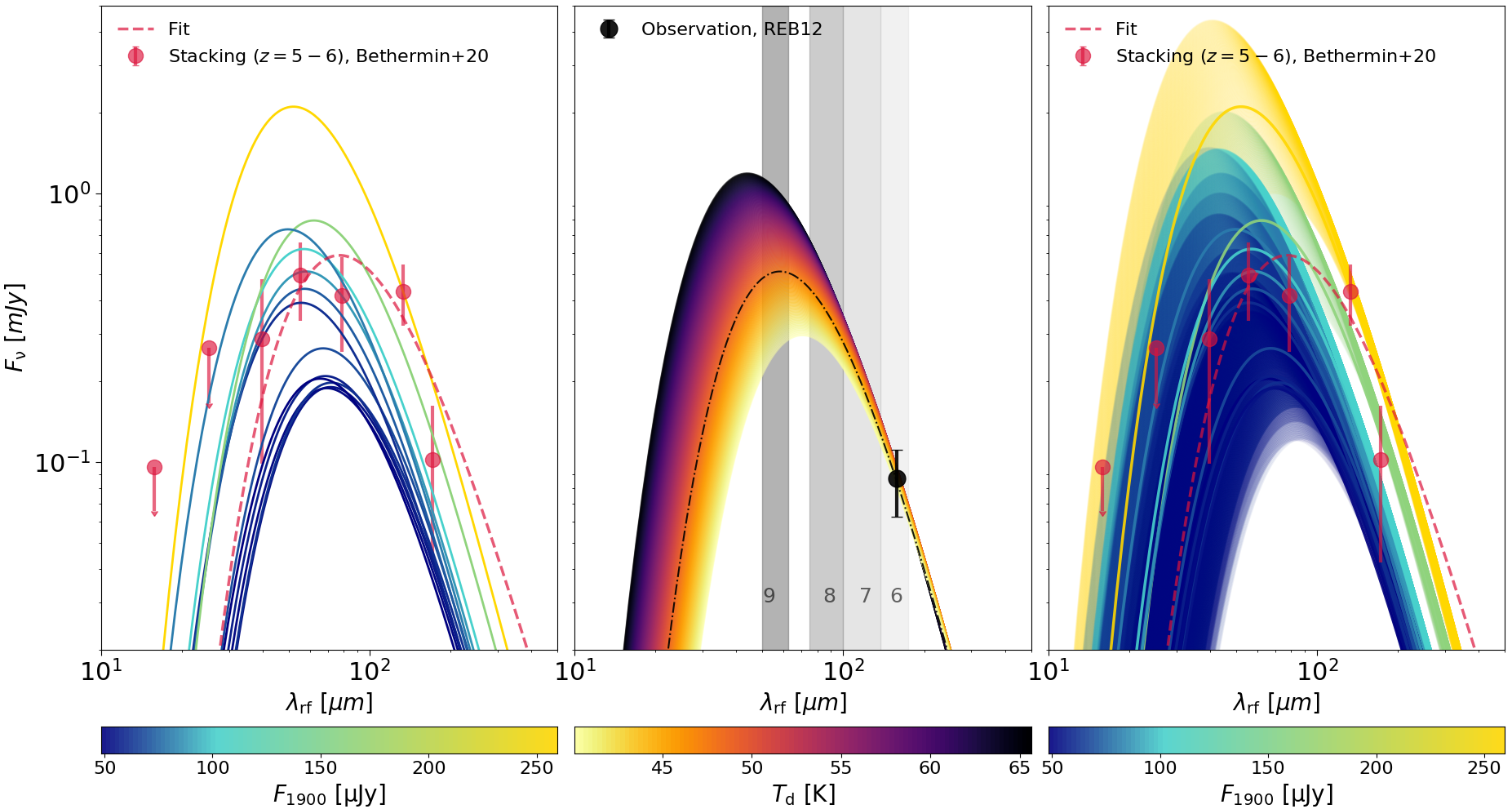}
    \caption{\textbf{Left panel:} FIR SEDs obtained using the median $(T_{\rm d}, M_{\rm d})$ derived for REBELS galaxies (see Table \ref{tabREB} for the sources details). The lines are colour-coded depending on their $1900\ \mathrm{GHz}$ flux (see colorbar). The red points are obtained by \citep{bethermin_alpine} for the ALPINE galaxies in the redshift interval $z=5-6$ through stacking. The red dashed line is the best fit to these points (corresponding to $T_{\rm d} \sim 42\ \mathrm{K}$);
    \textbf{Center:} Variation in the SED of a single source (REBELS-12) due to the $1-\sigma$ uncertainty in $(T_{\rm d}, M_{\rm d})$ around the median values. The lines are colour coded according to the corresponding dust temperature (see colorbar). The dashed black curve shows the SED for the median $T_{\rm d} \sim 46\ \mathrm{K}$. The black point is the continuum observation at $1900\ \mathrm{GHz}$. The shaded grey regions mark the ALMA bands $6$ to $9$.;
    \textbf{Right:} Variation of the SEDs of all the REBELS galaxies due to the uncertainty in $(T_{\rm d}, M_{\rm d})$ of each source. This plot can be interpreted as the combination of the left and central panels. Here we are showing the median FIR SEDs and their variation due to uncertainty on $(M_{\rm d}, T_{\rm d})$ for all the REBELS galaxies. The colour coding, the red points and red dashed line are the same as in the left panel.
    \label{SEDs}
    }
\end{figure*}

\section{Application to REBELS galaxies}\label{apllreb}

We use our method to derive $T_{\rm d}$ for the $13$ REBELS galaxies that are detected both in \CII\ and in the continuum at $1900\ \mathrm{GHz}$. In this work we always refer to this sub-sample of the survey targets. The observed properties of these sources are summarized in Table \ref{tabREB}. For the data analysis we refer to the dedicated papers by \citealt{inami:2022} (FIR continuum fluxes), \citealt{sander:2022} (\CII\ luminosities), and \citealt{Stefanon:2022} (stellar masses).

For some of the REBELS sources the UV half-light radius has been measured by \cite{bowlerREBradii}, who find on average $r_{\star} = 1.3 \pm 0.8\ \mathrm{kpc}$. \cite{bowlerREBradii} caution that the large sizes $r_{\star} > 1\ \mathrm{kpc}$ are due to multiple components, with individual clumps being less extended $r_{\star} \sim 0.2-1.0\ \mathrm{kpc}$. This is in agreement with the theoretical findings by \cite{ferrara:2022}, who predicts sub-kpc sizes for most REBELS galaxies. In light of this, we consider star-forming/dust emitting regions of REBELS galaxies to be uniformly randomly distributed in the range $0.2\ \mathrm{kpc}\simlt r_{\star} \simlt 1.1\ \mathrm{kpc}$. 

We then consider the \CII\ emitting regions to be $1.5 \simlt y \simlt 3$ times more extended than the UV emitting ones, as indicated by previous observations of galaxies at $z \simgt 4$ \citep[see e.g.][]{ carniani:2017oiii,carniani2018clumps,matthee2017alma,Matthee_2019,2019ApJ...887..107F,2020arXiv200300013F,2020A&A...633A..90G,Carniani20}.

The burstiness parameter of REBELS galaxies is unknown. High-$z$ UV selected sources are expected to be highly star forming and UV emitting by construction  \citep{2013MNRAS.434.1486D}. Both locally and at intermediate redshift, values up to $\kappa_{\rm s} \simeq 100$ have been observed in such galaxies \citep[see e.g.][]{Daddi_2010}. Recently, \cite{Vallini20} found for the mildly star-bursting COS-3018 at $z=6.854$ a value of $\kappa_{\rm s}\sim 3$, applying the \CII-emission model given in \cite{ferraraCII}. Applying the same method on $11$ bright UV selected galaxies\footnote{In particular: MACS1149-JD1 \citep{2018Natur.557..392H}, A2744-YD4 \citep{laporte:2017apj}, MACS416-
Y1 \citep{Tamura19,Bakx20}, SXDF-NB1006-2 \citep{2016Sci...352.1559I}, B14-65666 \citep{big3drag}, BDF3299 \citep{carniani:2017oiii}, J0121 J0235, J1211 \citep{2020Harik}. See also Table \ref{tabhz}, Fig. \ref{Tdz}.} at $z=6-9$, \cite{2021Vallini} find values as large as $10 \simlt \kappa_{\rm s} \simlt 80$. 
For the REBELS galaxies we choose a random uniform distribution in the range $1 \simlt \kappa_{\rm s} \simlt 50$ inline with the independent derivation in \citet{ferrara:2022} which suggests that these sources are not extreme starbursts. 

For the metallicity, which is also unconstrained by current data, we assume a broad uniform random distribution in the range $0.3-1\ \mathrm{Z_{\odot}}$. This general assumption is validated by numerical simulations of galaxies at $z\sim 6$ with similar stellar masses $10^9<M_{\star}<10^{11}$ \citep{Ma16, Torrey19}, and several observational studies which analyse FIR lines (such as \NII, \NIII, \CII, and \OIII) at $z \simgt 6-8$ to derive $Z$ \citep[see e.g.][and references therein]{sostpereira, big3drag,2019A&A...631A.167D,Tamura19,Vallini20,Bakx20,OIIImetal}. Current estimates of $Z$ at high redshift will be significantly ameliorated thanks to forthcoming ALMA observations and to the James Web Space Telescope (JWST) spectroscopy\footnote{JWST will provide observations of optical nebular lines (such as \Hb, \Ha, \NII, \OII\ and \OIII), which are reliable metallicity tracers, out to $z\sim 10$ \citep[see e.g.][]{wright2010tracing,Maiolino_2019,Chevallard}}.
 
We pause for a quick summary: the physical properties needed for the application of our method are $(\kappa_{\rm s},z,F_{\nu_ 0},Z,\Sigma_{\rm SFR},L_{\rm CII},y)$. For REBELS galaxies all these properties are constrained by observations, with the exception of $(y,\kappa_{\rm s},Z)$. For these three parameters, for each individual REBELS galaxy, we assume uniform random distributions in the broad ranges $1.5 \simlt y \simlt 3$, $1 \simlt \kappa_{\rm s} \simlt 50$, and  $0.3 < Z/Z_\odot < 1$, as motivated in the previous paragraphs.

Using eq. \ref{alfahz} we compute the \CII-to-total gas conversion coefficient $\alpha_{\rm CII}$ for all REBELS galaxies. Individual values are given in Table \ref{tabREB}, the average is $\left< \alpha_{\rm CII} \right>=5.3^{+0.8}_{-1.2}$ .
We note that this value is lower than the values we find in local galaxies (\citealt{TdCIImetodo}, $10 \simlt \alpha_{\rm CII} \simlt 10^{3}$ at $z\sim 0$). Interestingly, we find low $\alpha_{\rm CII} < 7$ also in the other $z>4$ sources analysed in \citealt{TdCIImetodo} (SPT0418-47 at $z=4.225$, B14-65666 at $z=7.15$, and MACS0416-Y1 at $z=8.31$). This might indicate that at fixed \CII\ luminosity high-$z$ galaxies have a lower gas content (for a detailed discussion see e.g. \citealt{ferraraCII}).

We conclude with a remark. The \textit{total} gas-to-\CII\ conversion factor computed here $\alpha_{\rm CII}=\Sigma_{\rm gas}/\Sigma_{\rm CII}$ is different by construction from the empirical \textit{molecular}-to-\CII\ conversion factor $\alpha_{\rm CII,mol}=\Sigma_{\rm H_2}/\Sigma_{\rm CII} = 31^{+31}_{-16}$ derived\footnote{In \citet{TdCIImetodo} we derive an analytical formula also for $\alpha_{\rm CII, mol} \approx 30.3\ t_{\rm depl, H_2}/\mathrm{Gyr}$, finding consistent values with \citet{zanita19} under the assumption of a molecular gas depletion time $t_{\rm dep, H_2} = 0.4-0.7\ \mathrm{Gyr}$ \citep{walter2020evolution}.} in \citet{zanita19}. The ratio of the two conversion factors corresponds to $\alpha_{\rm CII,mol}/\alpha_{\rm CII}=f_{\rm H_2}\ (r_{\rm gas}/r_{\rm H_2})^2 \approx f_{\rm H_2}\ y^2$ (under the reasonable assumption that $r_{\rm CII} \sim r_{\rm gas}$ and $r_{\rm H_2} \sim r_{\star}$), where $f_{\rm H_2}=M_{\rm H_2}/M_{\rm gas}$ is the molecular gas fraction. Since $f_{\rm H_2} < 1$ by definition, the empirical result $\alpha_{\rm CII,mol} \sim 31$ can be reconciled with the REBELS galaxies average value of $\left< \alpha_{\rm CII} \right> =5.3$ for $y>2.4$.

We proceed to compute $M_{\rm d}$ and $T_{\rm d}$ for all our targets. The results are summarised in Table \ref{tabREB}, and discussed in detailed in the following dedicated Sections.

\subsection{Dust temperatures}\label{TdREB}

We find the median $T_{\rm d}$ of REBELS galaxies to vary in the range between $39\ \mathrm{K}$ and $58\ \mathrm{K}$ with REBELS-39 (REBELS-18,32) being the galaxy with the warmest (coldest) dust. The median values are associated with large uncertainties (up to $\sim 35\%$) due to the lack of knowledge on the metallicity and burstiness of REBELS galaxies. Considering such uncertainties, the range widens to $30\ \mathrm{K}-73\ \mathrm{K}$ (where the lower limit is set by the condition on the dust masses and the upper one by that on the obscured SFRs, see  Section \ref{method}). Note that these large uncertainties on $T_{\rm d}$ are often comparable with the ones derived in the literature from traditional SED fitting using $2-3$ data points at FIR wavelengths (see Table \ref{tabhz} where we compare $T_{\rm d}$ values obtained from our method vs. SED fitting for several $z>5$ galaxies). Further ALMA observations in bands $8-9$ will help us significantly in constraining $T_{\rm d}$ at high-$z$ by sampling the SEDs closer to the FIR emission peak (see central panel of Fig. \ref{SEDs}). 

Overall we find an average dust temperature of $\left< T_{\rm d} \right>=47 \pm 6\ \mathrm{K}$. 
This value is in agreement with the independent derivation in \citet{ferrara:2022}, where they find $\left< T_{\rm d} \right> = (52 \pm 12)\ \mathrm{K}$. 
The physical model presented in \citet{ferrara:2022} relies on $1900\ \mathrm{GHz}$ continuum, and UV data (instead of \CII\ used here) to derive the physical properties of individual REBELS galaxies, such as their $(T_{\rm d}, M_{\rm d}, L_{\rm IR}, \mathrm{SFR})$, assuming assuming a given attenuation curve. 

The average value $\left< T_{\rm d} \right> \simeq 47\ \mathrm{K}$ is close to the dust temperature derived by \cite{bethermin_alpine} from the mean stacked SED of ALPINE sources (and analogs in the COSMOS field) in the redshift interval $z=5-6$ ($T_{\rm d} = 43 \pm 5\ \mathrm{K}$). For the lower redshift bin at $z=4-5$, where they have $3\times$ more sources (analogs included), \cite{bethermin_alpine} produce two different mean stacked SEDs dividing the considered sources depending on their SFR. For the galaxies with lower SFR (SFR $>10\ \mathrm{M_{\odot}/yr} $) they find $T_{\rm d}=41 \pm 1$ K, while for the higher star forming galaxies (SFR $>100\ \mathrm{M_{\odot}/yr}$) they find $T_{\rm d} = 47 \pm 2\ \mathrm{K}$. 

Since most REBELS galaxies have SFR $\simlt 100\ \mathrm{M_{\odot}/yr}$ (both from the DL relation, and from the independent analysis of rest-frame UV and IR data performed in \citealt{ferrara:2022}), they overall appear to have similar dust temperatures to the lower redshift ALPINE galaxies.  
We discuss this point in detail in the dedicated  Section \ref{tdredev}.

\subsection{Dust masses}\label{MdREB}

We find the median dust masses of REBELS sources to vary in the range $6.95 \leq \log (M_{\rm d}/M_{ \odot}) \leq 7.55$, REBELS-25 (REBELS-14,39) being the galaxy with the largest (lowest) dust mass. This is not surprising as REBELS-25 also has the largest stellar mass $M_{\star} \sim 10^{10}\ \mathrm{M_{\odot}}$ and continuum emission $F_{\rm 1900} = 260 \pm 22\ \mathrm{\mu Jy}$ in the sample. Overall we find an average dust mass around $\left< M_{\rm d} \right> =  (1.63\pm 0.73) \times 10^{7}\ \mathrm{M_{\odot}}$. This value is consistent with an independent derivation in \citet{ferrara:2022} ($\left< M_{\rm d} \right>= (1.3\pm 1.1) \times 10^7\ \mathrm{M_{\odot}}$), although they find a larger scatter in the dust masses values.  

We also compute the dust yield per SN, $y_{\rm d}$, required to produce REBELS galaxies dust masses. We use that: $y_{\rm d} = M_{\rm d}/M_{\star}\nu_{\rm SN}$, where $\nu_{\rm SN} = (53\ \mathrm{M_{\odot}})^{-1}$ is the number of SNe per solar mass of stars formed \citep{10.1046/j.1365-8711.2000.03209.x} assuming a Salpeter $1-100\ \mathrm{M_{\odot}}$ IMF.
For all (but two of) the REBELS sources we find $y_{\rm d} \simlt 1\ \mathrm{M_\odot}$ (see Table \ref{tabREB} for the value of $y_{\rm d}$ in each galaxy), which is quite consistent with the latest constraints on SNe dust production by \cite{lesniewska2019dust}. They find that up to $y_{\rm d} = 1.1\ \mathrm{M_{\odot}}$ of dust per SN can be produced, where the exact value depends on the amount of dust which is destroyed/ejected during the SN explosion ($1.1\ \mathrm{M_{\odot}}$ corresponds to the case of no dust destruction/ejection).   

We note that SN yield is still highly debated, with some works suggesting that dust destruction processes might only spare $\simlt 0.1\ \mathrm{M_{\odot}}$ per SN \citep[e.g.,][]{2016A&A...587A.157B,Matsuura2019,Slavin2020}. In this extreme case, inter-stellar medium grain growth or more exotic dust production mechanisms might be required \citep[][]{Mancini2015,Michalowski2015,2020MNRAS.494.1071G}. 

It is worth mentioning that the increase by a factor $\sim 3$ of the stellar masses possibly resulting from a non parametric SFH prescription would have a relevant impact in this context. In fact, as explained in Sec. \ref{method}, the dust masses remain mostly unaffected. This implies, by definition, a reduction by the same factor $\sim 3$ in the SN dust yield. As a result, we would find an average value of $\left< y_{\rm d} \right> \sim 0.23\ \mathrm{M_{\odot}}$ for the REBELS sources, consistent even with stringent SN dust production constraints.

For a detailed discussion we refer to \citet{dayal:2022} where they use the semi-analytical galaxy formation model DELPHI to investigate the dust build up and content of average $z \sim 7$ galaxies. The processes accounted for include SNe dust production, grain growth, astration, shock destruction, and ejection in outflows. The DELPHI model general predictions are then compared to the dust masses derived here for individual REBELS galaxies, finding a good agreement in most ($77\%$) cases. 

\subsection{IR emission}\label{IRREB}

In Fig. \ref{SEDs} we show the FIR SEDs obtained for the REBELS galaxies considering a single-temperature grey body approximation and the dust masses and temperatures in Table \ref{tabREB}. In the left panel we show the FIR SEDs obtained using the median $(T_{\rm d}, M_{\rm d})$ values of each source. We can see that the flux at the peak of emission changes within a factor $\times 10$, with a variation in the peak wavelength of emission $ 50\ \mathrm{\mu m} \leq \lambda_{\rm peak} \leq 73\ \mathrm{\mu m}$. 
Interestingly, almost all the $\lambda_{\rm peak}$ values that we predict lay within the range observable with ALMA, albeit in band-10 for the hottest sources. 

For the FIR SED of each source we refer to Appendix \ref{singleREB}. Here we focus on REBELS-12 (central panel of Fig. \ref{SEDs}), whose $M_{\rm d}$ and $T_{\rm d}$ are close to the average values in the sample. For REBELS-12 cold dust temperatures $T_{\rm d} < 30\ \mathrm{K}$ are disfavoured as they would result in very large dust masses $M_{\rm d} \gg 10^{7}\ \mathrm{M_{\rm \odot}}$, and very low $\kappa_{\rm s}$ values that are unexpected for a strongly UV-emitting high-z source. Very hot dust temperatures ($T_{\rm d} > 80\ \mathrm{K}$) are
similarly unlikely as they would result in very large IR luminosities, and consequently unreasonably large obscured SFRs (for REBELS-12 $\mathrm{SFR_{\rm IR}} > 500\ \mathrm{M_{\odot}/yr}$). These are hard to reconcile with the generally blue UV slopes $\beta$ observed in most REBELS galaxies (e.g. $\beta=-1.99$ for REBELS-12). We briefly discuss a possible caveat in  Section \ref{summary}.

\begin{figure}
    \centering
    \includegraphics[width=1.0\linewidth]{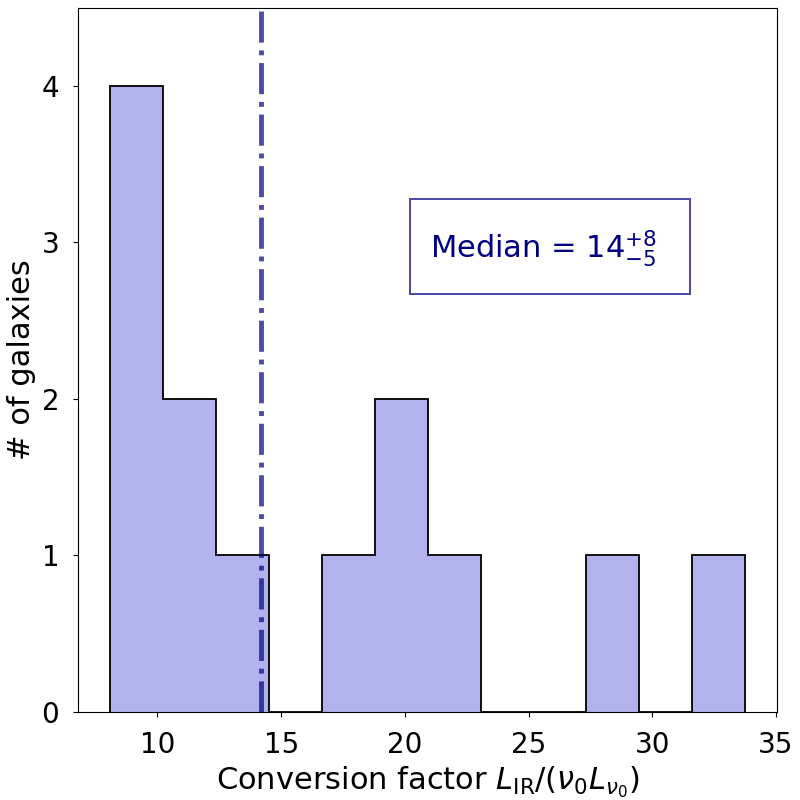}
    \caption{
    Conversion factor from the monochromatic luminosity $\nu_0 L_{\nu_0}$ at $1900\ \mathrm{GHz}$ to the total IR luminosity $L_{\rm IR}$ for the REBELS galaxies. We also show the median (blue line) value.}
    \label{convers}
\end{figure}

\subsubsection{$F_{\rm 1900}$ to total IR luminosity conversion at $z\sim 7$}

Using eq. \ref{LIR} we can compute the IR luminosities of all the REBELS galaxies. We find that their IR luminosities vary in the range  $1.7 \times 10^{11}\ \mathrm{L_{\odot}} \simlt L_{\rm IR}\simlt 2.8 \times 10^{12}\ \mathrm{L_{\odot}}$, which corresponds to obscured SFRs in the range $\sim 17-285\ \mathrm{M_{\odot}/yr}$ (assuming the conversion factor given in Table \ref{tabREB}). Interestingly, the most IR luminous REBELS (REBELS-25) classifies as an Ultra-Luminous InfraRed Galaxy (ULIRG, with $L_{\rm IR}>10^{12}\ \mathrm{L_{\odot}}$, see e.g. \citealt{Lonsdale06}) despite being UV selected. This peculiar source will be discussed in detail in \cite{Hygate:2022}.

Combining these values with the total star formation rates derived using the DL relation for starbursts we find that on average $\left< \mathrm{SFR}_{\rm IR}/\mathrm{SFR} \right> = 53\%$ (in agreement with the independent derivation in \citealt{2021arXiv210512133S,ferrara:2022,dayal:2022}), implying that REBELS galaxies are on average relatively UV-transparent sources. The only exceptions are represented by REBELS-8, REBELS-14, REBELS-25 and REBELS-39 for which we find $\mathrm{SFR_{\rm IR} \simgt SFR}$, i.e. the SFR deduced from \CII\ is exceeded (barely, only by $6\%$, in the case of REBELS-8). Interestingly these sources include $2$ (REBELS-8, and the peculiar REBELS-25) of the $4$ galaxies for which the UV-to-IR emission model in \citet{ferrara:2022} fails to find a solution for $(T_{\rm d},M_{\rm d})$ (using the same dust model adopted here, i.e. MW-like dust). In fact these galaxies have very large IR-to-UV flux ratios compared to their UV slopes. This might indicate that their UV and IR emitting regions are spatially segregated (see \citealt{ferrara:2022} for a detailed discussion).

We can provide a conversion factor from the galaxies monochromatic luminosity $\nu L_{\nu}$ at any frequency to their total $L_{\rm IR}$ (for a discussion on the application of such conversion factor see e.g. \citealt{2020MNRAS.491.4724F,Bouwens20}). The monochromatic luminosity at $1900\ \mathrm{GHz}$ can be written as:
\begin{equation}
    \nu_0 L_{\nu_0} =  2.92\times10^{11}\ \left[ \frac{g(6)}{g(z)} \right]\ \left( \frac{F_{\nu_0}}{\mathrm{mJy}} \right)\ L_{\odot}
\end{equation}
where $g(z)$ is defined in eq. \ref{defs}. The results for the conversion factor $L_{\rm IR}/\nu_0 L_{\nu_0}$ are shown in Fig. \ref{convers}. 

We find a median value of $L_{\rm IR}/\nu_0 L_{\nu_0}=14^{+8}_{-5}$ (where both $L_{\rm IR},\nu_0 L_{\nu_0}$ are in solar luminosity units). This range is consistent with the one obtained in \cite{Bouwens20} at $z\sim 7$ using a simple modified black body with $\beta_{\rm d}=1.6$, and a dust temperature derived from a linear fit of currently available dust temperature data vs. redshift. It is worth noting that they consider $T_{\rm d} = 54\ \mathrm{K}$ at $z=7$, which is warmer than the temperatures derived here for the REBELS galaxies. We will discuss this point in detail in the following  Section \ref{tdredev}.

\begin{figure*}
    \centering
    \includegraphics[width=1.0\linewidth]{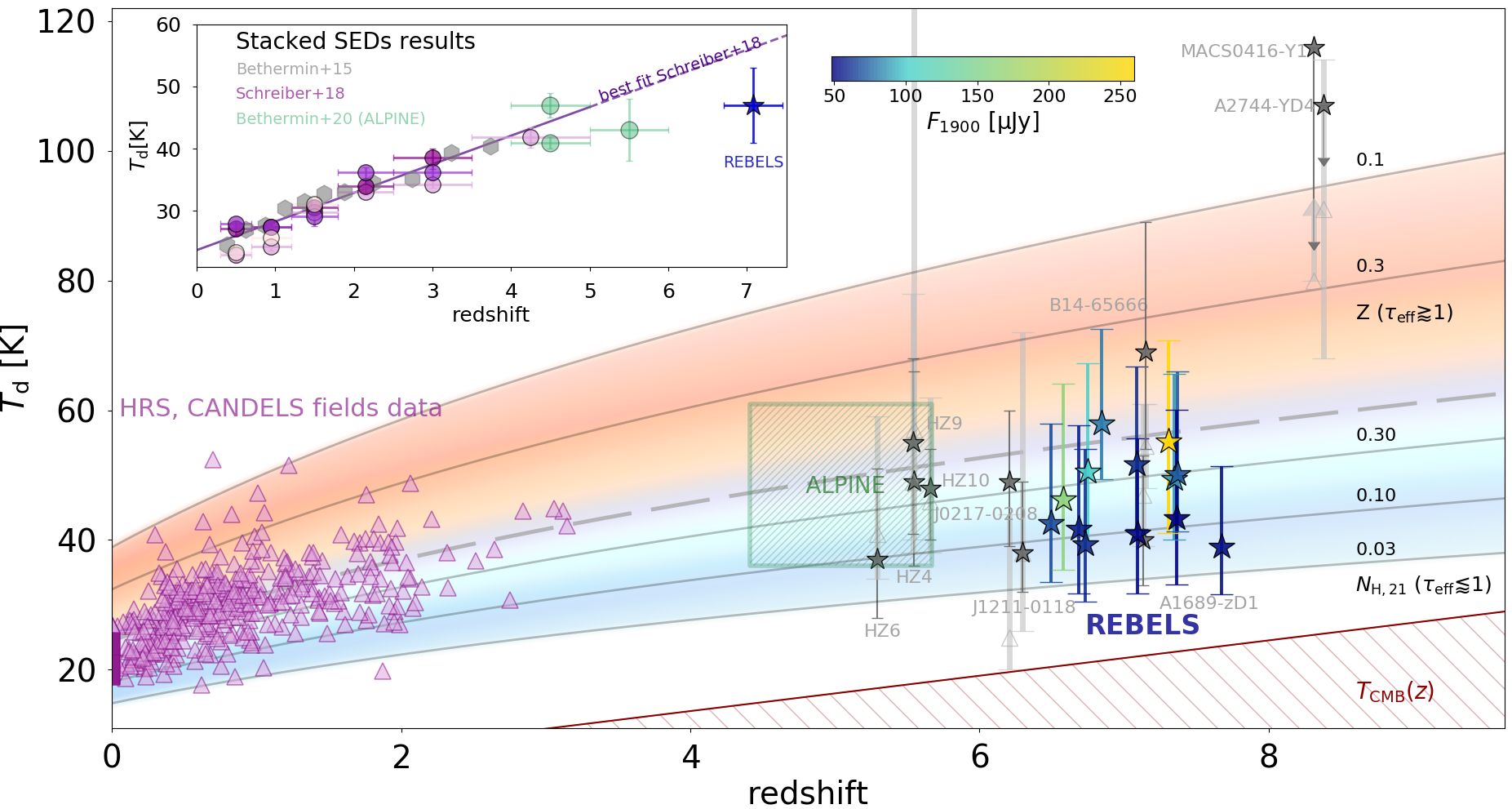}
    \caption{\textbf{Main panel}: We unpack the low-$z$ stacked SEDs into the single detections (purple triangles). These are the UV-to-IR normal star forming galaxies detected in the HRS and CANDELS fields \citep{Schreiber18}. We add the individual UV-selected galaxies at $z\simlt 5$ for which $T_{\rm d}$ estimates are available (see Table \ref{tabhz} for the sources details and references). We also show the $T_{\rm d}$ values obtained with our method for REBELS and ALPINE continuum detected galaxies (respectively, stars and green hatched rectangle). REBELS galaxies are colour-coded according to their $F_{1900}$ (see colorbar). The coloured region shows the $T_{\rm d}$-redshift evolution that we derive analytically in eq. \ref{Tdcosm}-\ref{approx} (for an increasing effective optical depth $\tau_{\rm eff}$ from blue to red). We find that on average $T_{\rm d}$ raises with redshift due to the decreasing gas depletion time at higher-$z$, as $T_{\rm d}(z) \propto t_{\rm dep}^{-1/6}$. The grey dashed line represents the relation given in eq. \ref{Tdcosm} which can be written explicitly as $T_{\rm d} = 24.5\ (1+1.5\ z)^{1/6}\ [ \Omega_{\rm m}(1+z)^3+\Omega_{\Lambda}]^{1/12}$ assuming $(\tau_{\rm eff},Z)$ equal unity. We can also explain the scatter in measured $T_{\rm d}$ at a given redshift. In the UV-transparent approximation ($\tau_{\rm eff} \lessapprox 1$) the scatter depends solely on the column density $N_{\rm H,21}=N_{\rm H}/10^{21}\ \mathrm{cm}^{-2}$ (see solid lines and associated labels). In the UV-obscured approximation ($\tau_{\rm eff} \gtrapprox 1$) the scatter depends only on the metallicity $Z$ (see solid lines and associated labels; note that $Z$ is in solar units). \textbf{Inset panel}: $T_{\rm d}$ values obtained from stacked SEDs fitting in the redshift range $0\simlt z \simlt 6$ (grey, purple and green points, from \citealt{2015Bet,Schreiber18,bethermin_alpine}, respectively). The purple line shows the linear best fit obtained by \citet{Schreiber18} at $z \simlt 4$. The blue star corresponds to the average temperature $T_{\rm d}=(47 \pm 6)\ \mathrm{K}$ derived here for the REBELS galaxies. 
    }
    \label{Tdz}
\end{figure*}

\section{Dust temperature evolution}\label{tdredev}

In the last few years, several works have suggested the presence of a trend of increasing dust temperature with redshift in star forming galaxies detected in the UV-to-IR rest-frame wavelength range \citep[][]{2012ApJ...760....6M,2013Magnelli,2014A&A...561A..86M,2015Bet,Schreiber18}. 

In particular, \cite{Schreiber18} fitted stacked SEDs to a complete sample of main sequence galaxies in the CANDELS fields, and derived the linear relation $T_{\rm d} = 32.9 + 4.6\,(z-2)$, shown in the inset of Fig. \ref{Tdz}.  Compared to those for individual sources, stacked SEDs (a) reduce the scatter introduced in the relation by different intrinsic galaxy properties, and (b) allow to extend the relation up to $z \sim 4$. The above trend is also consistent with the stacked SED fitting results produced for ALPINE galaxies by \cite{bethermin_alpine} at $4<z<6$ (also shown in the inset). In this case, a stacking procedure was necessary since ALPINE galaxies are individually detected only in a single band at restframe $1900\ \mathrm{GHz}$. Hence it would be impossible to constrain their $T_{\rm d}$ with ordinary SED-fitting.

More recently, $T_{\rm d}$ obtained from stacked SEDs at $z<6$ has been combined with the $T_{\rm d}$ derived for individual galaxies detected in multiple bands at $z \simgt 5$. Two studies reached somewhat discordant conclusions. While \cite{Liang19,dusttemp2020} deduce a flattening of $T_{\rm d}-z$ relation at $z>4$, \cite{Bouwens20} confirm the trend with little modification in the slope. 

Either trends are yet to be explained by a physical model. Previous theoretical works suggested that warmer dust temperatures at high-$z$ might depend on the dust geometry, with most of the dust in high-$z$ galaxies being located in compact, young star-forming regions, where the dust heating is particularly efficient \citep{behrens18,Liang19,sommovigo20}. Other works \citep{2014A&A...561A..86M,Ma16,2019MNRAS.487.1844M,2021arXiv210412788S} have suggested that the increasing $T_{\rm d}(z)$ trend together with the flattening at $z\sim 4$ are correlated to the specific star formation rate increase with redshift and its subsequent plateau at $z \sim 4$ \citep[see e.g.][]{Tomczak_2016,Santini_2017}. Finally the lower dust-to-metal ratio predicted for early galaxies from simulations at $z \simgt 5$ \citep[see e.g.][]{2017MNRAS.466..105A,behrens18,2021arXiv210412788S,pallottini:2022} could also motivate the presence of hot dust at high-$z$. In fact, a lower dust content results in warmer dust temperatures for a given $L_{\rm FIR}$.

Our analysis of the REBELS sample can clarify the issue of the $T_{\rm d}(z)$ evolution in a unique way for two reasons: (a) REBELS continuum detections double the number of previously known sources at $z \simgt 5$; (b) possible deviations from the linear trend are more robustly identified at the higher ($z\approx 7$) redshift of the REBELS sample.

Extrapolating the \cite{Schreiber18} relation to $z= 7$ gives $T_{\rm d}= (56 \pm 4)\ \mathrm{K}$, i.e. a slightly warmer temperature than found here, $T_{\rm d} \sim  47\ \mathrm{K}$. This does not come as a surprise, as there is no special physical motivation to expect a linearly increasing $T_{\rm d}$ trend. It is therefore crucial to develop a simple but physical theoretical framework against which observations of individual UV-to-FIR detected galaxies in the range $0 \simlt z \simlt 8.5$ can be compared and interpreted. This is our goal in the next Section.

\subsection{Physical origin of the $T_{\rm d}-z$ relation}\label{teorTd-z}

Under the assumption that FIR and UV emitting regions are co-spatial, $L_{\rm IR}$ is equal to the absorbed fraction of the intrinsic UV luminosity, $L_{\rm 1500}$:
\begin{equation}\label{lfir1}
    L_{\rm IR} = (1-e^{-\tau_{\rm eff}}) L_{\rm 1500} = (1-e^{-\tau_{\rm eff}}) {\cal K}_{\rm 1500} \mathrm{SFR}.
\end{equation}
In the previous equation, $\tau_{\rm eff}$ is the galaxy effective dust attenuation optical depth at $1500\ \angstrom$, which directly relates to the transmissivity, $\tau_{\rm eff}=-\ln\,T$. The transmissivity is defined as the ratio of the observed-to-intrinsic UV luminosity, i.e. $T = 1$ ($T = 0$) for a fully transparent (obscured) galaxy. Note that, depending on the relative dust and star distributions, $\tau_{\rm eff}$ might significantly differ from the physical UV optical depth, $\tau_{1500}$. Hence, a large transmissivity does not directly imply a low physical optical depth. The conversion factor for (a) continuous star formation at a fix age of 150 Myr, (b) Salpeter IMF in the range 1-100 $M_\odot$, (c) metallicity $Z=1/3\ \mathrm{Z_{\odot}}$ is ${\cal K}_{\rm 1500} \equiv L_{\rm 1500}/{\rm SFR} =1.174 \times 10^{10}\ {L_\odot}/(M_\odot {\rm yr}^{-1})$ \citep{ferrara:2022}.

Combining eq. \ref{LIR} and \ref{lfir1}, and recalling that $M_{\rm d} = D M_{\rm g}$, it follows that 
\begin{equation}\label{Tdcosm}
    T_{\rm d} = 29.7 \left[ \frac{(1-e^{-\tau_{\rm eff}})}{Z}\left(\frac{\rm Gyr}{t_{\rm dep}}\right) \right]^{1/(4+\beta_d)} {\rm K}
\end{equation}
where $\beta_{\rm d}=2.03$ and $Z$ is in solar units. We have introduced the \textit{total} gas depletion time $t_{\rm dep} = M_{\rm g}/{\rm SFR}$, which we derive in the following from cosmological arguments. Eq. \ref{Tdcosm} shows that the dust temperature is larger in optically thick, low metallicity systems with a short depletion time. 

Let us express $M_{\rm g}$ and SFR of a galaxy in terms of its total (dark + baryonic) halo mass, $M$, and mean dark matter accretion rate, $\langle dM/dt \rangle$:
\begin{align}
        M_{\rm g} & = f_{\rm b} M -M_{\star} \label{Mgsist}\\
        \mathrm{SFR} & = \epsilon_{\star} f_{\rm b}  \left< \frac{dM}{dt} \right> \label{sfrsist}
\end{align}
where $\epsilon_{\star}$ is the star formation efficiency, and $f_{\rm b}=\Omega_{\rm B}/\Omega_{\rm M}$ is the (cosmological) baryon fraction in the halo. By integrating eq. \ref{sfrsist} and substituting it into eq. \ref{Mgsist} we obtain
\begin{equation}\label{Mg}
    M_{\rm g}=f_{\rm b} M (1-\epsilon_{\star})
\end{equation}

Numerical simulations \citep{Fakhouri10,Dekel13,2015DM} provide the following fit to the mean halo accretion rate, as a function of redshift and halo mass, $M_{12}=M/10^{12} M_\odot$,
\begin{equation}\label{dmdt}
    \left< \frac{dM}{dt} \right> = 69.3\ M_{\rm 12}\ f'(z)\ E(z)\;\  M_{\odot}{\rm yr}^{-1},
\end{equation}
with  
\begin{equation}\label{funct_dmdt}
    f'(z) =-0.24+0.75(1+z); \quad E(z) = [\Omega_{m}(1+z)^3 + \Omega_{\Lambda}]^{1/2}.
\end{equation}
By combining the previous equations $t_{\rm dep}$ takes the form 
\begin{equation}\label{tdepl}
    t_{\rm dep}(z) =  {t_{\rm d,0}}{[f'(z)E(z)]^{-1}}, 
\end{equation}
where the timescale $t_{\rm d,0} = 14.4 (1-\epsilon_{\star})/\epsilon_{\star} \,{\rm Gyr}$ is fixed so that $t_{\rm dep}\ (z=0)= 2\ \mathrm{Gyr}$ as approximately measured in local galaxies, including the MW \citep[][]{2008AJ....136.2846B,Leroy_2008,2010MNRAS.407.2091G}.  

According to eq. \ref{tdepl}, the depletion time decreases with redshift as $(1+z)^{-5/2}$ as a result of the higher cosmological accretion rate at early times\footnote{While $t_{\rm dep}$ is fundamentally unknown at high-$z$, the \textit{molecular} gas depletion time, $t_{\rm dep,H_2}=M_{\rm H_2}/\mathrm{SFR}$, has been studied up to $z\simlt 5$ via CO and dust observations \citep[see e.g.][]{2018ApJ...853..179T,walter2020evolution,2020ARA&A..58..157T}. For instance,  \cite{2020ARA&A..58..157T} suggest a mild evolution of $t_{\rm dep,H_2}$ time with redshift for main-sequence galaxies, $t_{\rm dep,H_2} \propto (1+z)^{-0.98\pm 0.1}$. Assuming such $t_{\rm dep,H_2}(z)$ evolution (instead of the one given in eq. \ref{tdepl}), does not qualitatively affect our results. In fact, $T_{\rm d}$ would still increase with redshift due to the shorter depletion times. However, the different evolution slightly modifies the predicted $(N_{\rm H,21}, Z)$ values shown in the Figure.}. This point is crucial as, barring variations of the optical depth and metallicity (see below), the redshift evolution of $T_{\rm d}$ is governed by the gas depletion time in galaxies. From the result above, and using eq. \ref{Tdcosm}, it follows that 
\begin{equation}\label{approxTcosm}
T_{\rm d} \propto (1+z)^{5/2(4+\beta_d)} \approx (1+z)^{0.42}.
\end{equation}
As we will see shortly, this trend matches perfectly the observed one. 

On top of the above overall increasing trend of dust temperature, at fixed redshift scatter is introduced by variations of $\tau_{\rm eff}$ and $Z$ in individual galaxies. It is useful to separately discuss two asymptotic regimes, i.e. the optically thin ($1-e^{-\tau_{\rm eff}} \approx \tau_{\rm eff}$), and optically thick ($1-e^{-\tau_{\rm eff}} \approx 1$) one. In these two limits eq. \ref{Tdcosm} becomes
\begin{equation}\label{approx}
 \left\{\begin{aligned}
    \ T_{\rm d} = &\  29.4\ \left[ N_{\rm H,21}\ f'(z)\ E(z) \right]^{1/6.03} {\rm K}        \quad & \tau_{\rm eff} \lessapprox 1\; (T \gtrapprox 37\%),\\ 
    \ T_{\rm d} = &\ 29.6\ \left[ \frac{f'(z)\ E(z)}{Z} \right]^{1/6.03}  {\rm K} \quad & \tau_{\rm eff} \gtrapprox 1\; (T \lessapprox 37\%); 
    \end{aligned}\right.
\end{equation}
we have used eq. \ref{DtoZ} to write $\tau_{\rm eff}/Z = \sigma_{\rm ext}N_{\rm H}$, where $N_{\rm H}= 10^{21}\  N_{\rm H,21}\ {\rm cm}^{-2}$ is the effective gas column density\footnote{As in general, $\tau_{\rm eff} < \tau_{1500}$, $N_{\rm H}$ should be intended as a lower limit to the actual mean column density in the galaxy.} of the galaxy, and $\sigma_{\rm ext}=0.96\times 10^{-21} {\rm cm}^{2}$ is the extinction cross-section appropriate for the adopted dust model. Eq. \ref{approx} is graphically displayed in Fig. \ref{Tdz}, where it is also compared with available data. 

Interpreting eq. \ref{approx} is straightforward. First, for optically thin galaxies (for which UV transmissivity $T \gtrapprox 37\%$) $T_{\rm d}$ depends solely on $N_{\rm H}$, with larger column densities producing warmer dust. Quantitatively, for a $z=0$ source with $N_{\rm H,21} =1$, corresponding to a ratio $\tau_{\rm eff}/Z = 0.96$, we predict $T_{\rm d} = 26.5\ \mathrm{K}$. Second, dust in obscured sources (for which $T \lessapprox 37\%$) is warmer; this is not surprising, as a larger obscuration results in more efficient dust heating. Third, among obscured sources, $T_{\rm d}$ is higher in metal-poor systems. This is because a lower metallicity implies a smaller dust content, which for fixed $L_{\rm IR}$ results in warmer temperatures. 

Locally, given the observed ($N_{\rm H}, Z$) scatter in individual sources, our model predicts variations as large as $\Delta T_{\rm d} \simeq 25\ \mathrm{K}$. We finally note that, if the ($N_{\rm H}, Z$) range does not evolve with time, the scatter at $z=0$ gets amplified at earlier times by the redshift dependence of $T_{\rm d}$, reaching  $\Delta T_{\rm d} \simeq 55\ \mathrm{K}$ at $z=8$.

\subsection{Comparison with observations at $0 \simlt z \simlt 8$}\label{sec:td_comparison}

We now intend to compare our theoretical predictions with dust temperature estimates available in the literature for UV-detected sources at $0 \simlt z \simlt 8$. 

We recover the dust temperatures of individual UV-to-IR detected galaxies whose stacked SEDs in the redshift range $0 \simlt z \simlt 3$ are used in the analysis by \cite{Schreiber18}. We then add all the UV-selected galaxies at $z \simgt 5$ for which dust temperature estimates are available in the literature (see Table \ref{tabhz} for details of the sources). We apply the method used here to derive $T_{\rm d}$ for these galaxies, finding values consistent (within $1-\sigma$) with SED fitting results (see Table \ref{tabhz} for the detailed comparison). Finally, we apply our method to individual ALPINE galaxies detected simultaneously in \CII\ and continuum. We find their median dust temperatures to vary in the range $35\ \mathrm{K} \simlt T_{\rm d} \simlt 60\ \mathrm{K}$, which is consistent with the stacked SEDs fitting results in \cite{bethermin_alpine} ($40\ \mathrm{K} \leq T_{\rm d} \leq 49\ \mathrm{K}$). A detailed analysis of ALPINE galaxies will be presented in \citet{sommovigo:2022}. 

The complete collection of $T_{\rm d}$ values is shown in Fig. \ref{Tdz} as a function of redshift. We stress that we consistently compare dust temperatures obtained by fitting individual galaxy SEDs; moreover, the same method is applied to \textit{all} high-$z$ sources ( see Appendix \ref{highzappl}). This avoids the confusion arising from comparing intrinsically different quantities such as dust temperatures obtained from stacked SEDs, and/or peak dust temperatures $T_{\rm peak} \sim  2.9\times 10^3 (\lambda_{\rm peak}/\mathrm{\mu m})^{-1}$.

The physical interpretation of $T_{\rm peak}$ might be unclear for $z>5$ galaxies. Indeed currently available data at these redshifts hardly trace the peak of FIR emission. Moreover, when a different SED fitting function other than the optically thin grey-body is used, $T_{\rm peak}$ can significantly differ from $T_{\rm d}$. In fact the assumptions made for the MIR (rest-frame) portion of the spectra affect $T_{\rm peak}$ \citep{dusttemp2020}, and the validity of such assumptions cannot be tested as no currently available instrument probes MIR wavelengths at $z>5$.

We find that our predictions are in agreement with data. Fitting all the dust temperatures with a single power law: $T_{\rm d}(z) = az^{\alpha}+b$, we find $\alpha=(0.58 \pm 0.04)$, which is close to the value $0.42$ given in eq. \ref{approxTcosm}. The slight difference is due to the fact that $T_{\rm d}$ does not depend uniquely on redshift, as discussed in detail in the previous section (see eq. \ref{approx}). Hence fitting all the data with a single power-law is misleading. 

The additional dependence on the column density (for optically thin sources), and metallicity (for optically thick sources) is responsible for the scatter in the measured temperatures at a given redshift. At $z \simeq 0.3$ variations as large as $\Delta T_{\rm d} = 22\ \mathrm{K}$ are observed \citep{Schreiber18}, which is perfectly consistent with our predictions ($\Delta T_{\rm d} \simeq 25\ \mathrm{K}$ in the local Universe). 

The amplification of the dust temperature scatter at high-$z$ that we predict (if the $N_{\rm H},Z$ range  does not evolve) is also consistent with data. In the narrow redshift range $7.6 \simlt z \simlt 8.3$ variations as large as $\Delta T_{\rm d} = 53\ \mathrm{K}$ are observed (we predicted $\Delta T_{\rm d} \simeq 55\ \mathrm{K}$ at $z =8$). 

At one extreme there are galaxies hosting very hot dust such as MACS0416-Y1 and A2744-YD4 ($T_{\rm d} > 80\ \mathrm{K}$ \citealt{Bakx20}, and $T_{\rm d}=90\ \pm 20\ \mathrm{K}$ \citealt{laporte:2017apj,behrens18}, respectively). On the other, there are galaxies such as REBELS-18 and J1211-0118 that show more moderate dust temperatures, possibly closer to local sources ($T_{\rm d}= 39^{+12}_{-7}\ \mathrm{K}$ and $T_{\rm d}=38^{+16}_{-8}$ \citealt{inoue20}, respectively).

Our physical model suggests that MACS0416-Y1 and A2744-YD4 are more heavily obscured than most currently observed high-$z$ galaxies. Larger $\tau_{\rm eff}>1$ combined with low metallicities $Z/Z_{\odot} \simlt 0.3$ can explain the very hot dust temperatures found in these two galaxies. 

We explain the colder dust temperatures found in REBELS-18 and J1211-0118 with lower effective optical depths $\tau_{\rm eff}<1$, i.e. larger UV transmissivity (ultimately resulting in less efficient dust heating). For these galaxies we predict mean gas column densities around $0.3\  \times 10^{20}\ \mathrm{cm^{-2}} \simlt N_{\rm H} \simlt 1.0\ \times 10^{20}\ \mathrm{cm^{-2}}$.   

\section{Summary and conclusions}\label{summary}

We have applied a novel method \citep[][]{TdCIImetodo} to derive the dust temperature to 13 $z \approx 7$ galaxies part of the ALMA Large Program REBELS. Our method combines the continuum and super-imposed \CII\ line emission measurements, thus breaking the SED fitting degeneracy between dust mass and temperature. This allows us to constrain $T_{\rm d}$ from a single-band restframe observation at $1900\ \mathrm{GHz}$, and to derive dust masses, IR luminosities, and the obscured SFR. Moreover, since REBELS targets constitute the first significant sample of continuum detected sources at $z\sim 7$ (for which $T_{\rm d}$ estimates are available), we can extend the reported $T_{\rm d}$-redshift relation \citep{2012Magdis,2013Magnelli,2015Bet,Schreiber18,dusttemp2020,Bouwens20} into the Epoch of Reionization. 

We summarize below our main findings:
\begin{itemize}
    \item \textbf{Dust temperature and mass}: the median $T_{\rm d}$ values for REBELS galaxies vary in the range $39-58\ \mathrm{K}$, with $\sim 35\%$ associated  uncertainty. The median dust masses are in the narrow range $(0.9-3.6)\times 10^7 M_\odot$. Dust production from SNe alone in most cases ($85\%$) can generate such dust masses assuming a dust yield $\simlt 1\ \mathrm{M_{\odot}}$ per SN; 
    
    \item \textbf{IR luminosities and $L_{\rm 1900}$-to-$L_{\rm IR}$ conversion at $z\sim 7$}: REBELS galaxies IR luminosities vary in the range  $1.7 \times 10^{11}\ \mathrm{L_{\odot}} \simlt L_{\rm IR}\simlt 2.8 \times 10^{12}\ \mathrm{L_{\odot}}$, which corresponds to obscured SFRs around $\sim 17-285\ \mathrm{M_{\odot}/yr}$ . We also derive their average conversion factor $L_{\rm IR} = 14^{+8}_{-5}\, L_{\rm 1900}$, where  $L_{\rm 1900} = \nu_0 L_{\nu_0}$, with $\nu_0=1900\ \mathrm{GHz}$. This value is consistent with an extrapolation of the empirical fitting formula of \citet{Bouwens20} to $z \approx 7$;
    
    \item \textbf{Dust temperature cosmic evolution}: we produce a \textit{new physical model} (see eq. \ref{Tdcosm}-\ref{approx}) that motivates the dust temperature increase with redshift. Such trend is an imprint of the decreasing gas depletion time towards high-$z$, $t_{\rm dep} \propto (1+z)^{5/2}$. We show that $T_{\rm d} \propto t_{\rm dep}^{-1/6}$, or $T_{\rm d} \approx (1+z)^{0.42}$; 
    
    \item \textbf{Dust temperature scatter at a given redshift}: on top of the $T_{\rm d}-z$ trend, we can also physically motivate the scatter in the measured $T_{\rm d}$ values at a given redshift. We find that in UV-transparent galaxies (UV transmissivity $\gtrapprox 37\%$) the scatter in $T_{\rm d}$ depends solely on the column density $N_{\rm H}$, with larger $N_{\rm H}$ corresponding to hotter dust. Instead, in UV-obscured galaxies the scatter in $T_{\rm d}$ depends only on the metallicity $Z$, with lower $Z$ implying hotter dust.
\end{itemize}

A very hot dust component, implying a large obscured SFR, can coexist with a steep UV slope in the presence of \textit{spatial segregation} of IR and UV emitting regions. This possibility has been suggested by theoretical studies and simulations in some $z \sim 7-8$ sources \citep{behrens18,Liang19,sommovigo20}. Such scenario is also supported by some observations; for instance \cite{2012Hodge,2016Hodge,carniani:2017oiii,laporte:2017apj,bowler2018obscured,2021MNRAS.tmp.3413B} find significant spatial offset between their ALMA and HST data. High-resolution ALMA follow-up observations of REBELS galaxies are required in order to make a step forward (see also \citealt{inami:2022,ferrara:2022} for a discussion on this point). 

JWST will also provide us with much more accurate metallicity measurements at $z \simgt 5$, improving current estimates of the dust-to-gas ratios at high-$z$. Finally, further ALMA observation at shorter wavelengths in band $7-8-9$, will allow us to reduce the uncertainties in current dust temperatures estimates at $z \simgt 5$ by sampling galaxies SEDs closer to the FIR emission peak \citep[see Fig. \ref{SEDs} and the discussion in ][]{bakx:2021}.

\section*{Acknowledgements}
LS, AF, AP acknowledge support from the ERC Advanced Grant INTERSTELLAR H2020/740120 (PI: Ferrara). Any dissemination of results must indicate that it reflects only the author’s view and that the Commission is not responsible for any use that may be made of the information it contains. Partial support from the Carl Friedrich von Siemens-Forschungspreis der Alexander von Humboldt-Stiftung Research Award is kindly acknowledged (AF). PD acknowledges support from the ERC starting grant DELPHI StG-717001, from the NWO grant ODIN 016.VIDI.189.162 and the European Commission's and University of Groningen's CO-FUND Rosalind Franklin program. RJB and MS acknowledge support from TOP grant TOP1.16.057.  SS acknowledges support from the Nederlandse Onderzoekschool voor Astronomie (NOVA). RS and RAB acknowledge support from STFC Ernest Rutherford Fellowships [grant numbers ST/S004831/1 and ST/T003596/1].  RE acknowledges funding from JWST/NIRCam contract to the University of Arizona, NAS5-02015.  PAO, LB, and YF acknowledge support from the Swiss National Science Foundation through the SNSF Professorship grant 190079 \quotes{Galaxy Build-up at Cosmic Dawn}. HI and HSBA acknowledge support from the NAOJ ALMA Scientific Research
Grant Code 2021-19A. HI acknowledges support from the JSPS KAKENHI Grant Number JP19K23462. JH gratefully acknowledges support of the VIDI research program with project number 639.042.611, which is (partly) financed by the Netherlands Organisation for Scientific Research (NWO). MA acknowledges support from FONDECYT grant 1211951, \quotes{ANID+PCI+INSTITUTO MAX PLANCK DE ASTRONOMIA MPG 190030}, \quotes{ANID+PCI+REDES 190194} and ANID BASAL project FB210003. LG and RS acknowledge support from the Amaldi Research Center funded by the MIUR program \quotes{Dipartimento di Eccellenza} (CUP:B81I18001170001). YF further acknowledges support from NAOJ ALMA Scientific Research Grant number 2020-16B \quotes{ALMA HzFINEST: High-z Far-Infrared Nebular Emission STudies}. IDL acknowledges support from ERC starting grant DustOrigin 851622.  JW acknowledges support from the ERC Advanced Grant QUENCH 695671, and from the Fondation MERAC. EdC gratefully acknowledges support from the Australian Research Council Centre of Excellence for All Sky Astrophysics in 3 Dimensions (ASTRO 3D), through project number CE170100013.

\section*{Data Availability}
Data generated in this research will be shared on reasonable request to the corresponding author.

\bibliographystyle{stile/mnras}
\bibliography{bibliografia/bibliogr,bibliografia/codes}

\newpage
\appendix{}

\section{Individual SEDs}\label{singleREB}

\begin{figure*}
\centering
\includegraphics[width=0.85\textwidth]{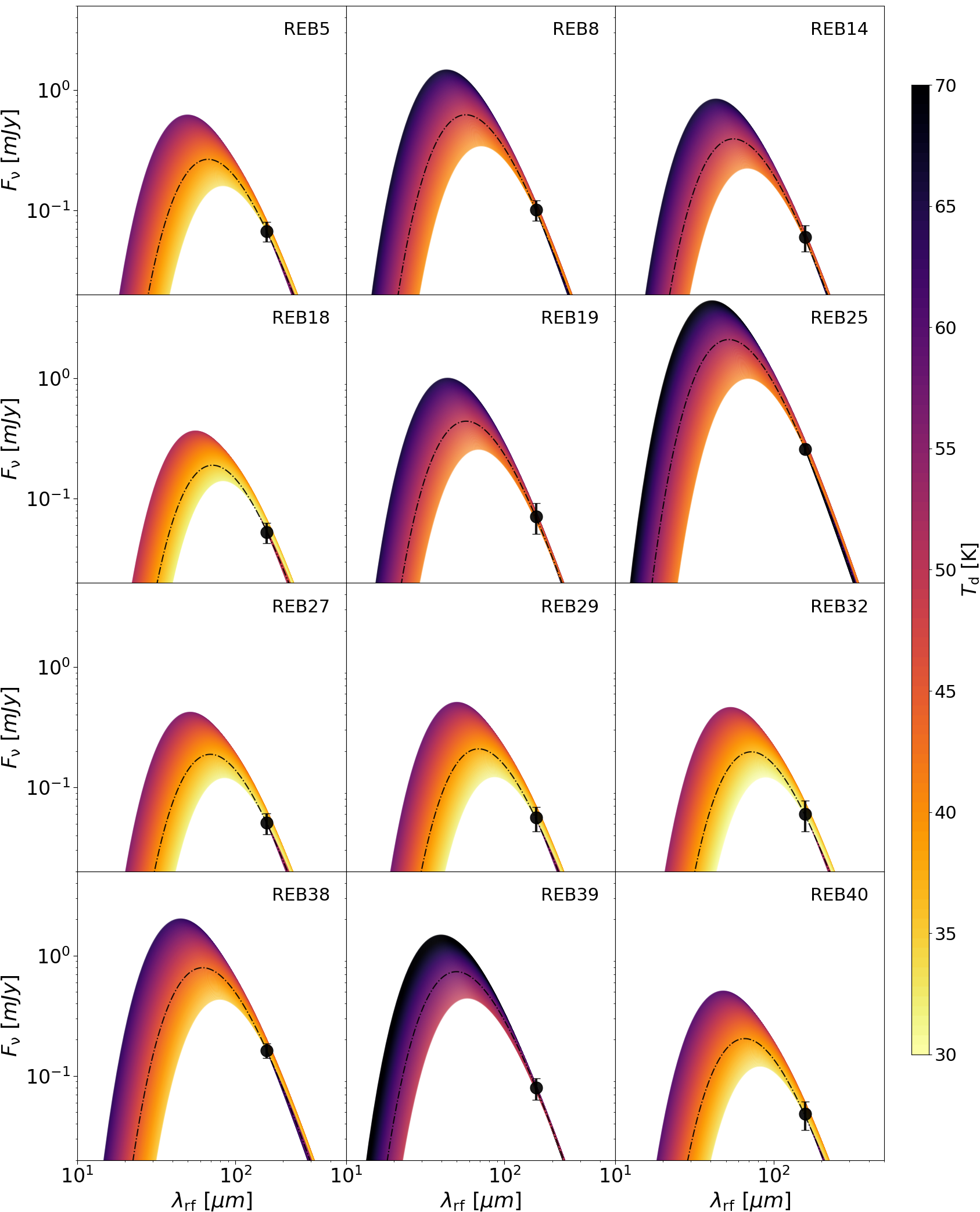}
\caption[short]{ Variation in the derived SEDs of all REBELS \CII\ and continuum detected galaxies due to the $1-\sigma$ uncertainties in their individual $(T_{\rm d}, M_{\rm d})$. The SEDs are colour coded according to the corresponding dust temperatures (see colorbar). The dashed black curves show the SEDs obtained with the median ($T_{\rm d},M_{\rm d}$) values for each galaxy. The black points represent the continuum observations at $1900\ \mathrm{GHz}$. For further details on the sources see Table \ref{tabREB}).}
\label{indivSEDS}
\end{figure*}

In Fig. \ref{indivSEDS} we show the individual FIR SEDs obtained for all REBELS \CII\ and continuum detected galaxies (analogously to the central panel in Fig. \ref{SEDs} representing REBELS-12 only). For all the sources we show the SEDs obtained with their median values of $(T_{\rm d}, M_{\rm d})$, and the variation due to the 1-$\sigma$ uncertainties in these two quantities (on average $\Delta T_{\rm d}/T_{\rm d} \sim 30 \% $ and $\Delta M_{\rm d}/M_{\rm d} = 70 \% $). 

These large uncertainties result from the lack of information on the metallicity $Z$ and burstiness parameter $\kappa_{\rm s}$ of REBELS galaxies (see the discussion in  Section \ref{apllreb}). Future observations will help us to constrain both these quantities, thus reducing the uncertainties in the predicted $T_{\rm d},M_{\rm d}$. In fact, with ALMA we can investigate the \OIII/\CII\ luminosities ratios of REBELS sources, which can be used to reliably constrain their $\kappa_{\rm s}$ using the model in \citealt{2021Vallini} (these ALMA observations would also provide us with an additional data point in the FIR continuum underlying \OIII). Moreover, future JWST optical nebular lines observations will allow us to improve metallicity estimates out to very high-$z$, possibly reaching a precision as low as $\Delta Z/ Z \sim  35 \%$ at $z \sim 7$ \citep{wright2010tracing,Chevallard,Maiolino_2019}.

\section{Application to other samples}\label{highzappl}

In Table \ref{tabhz} we show the comparison between measured dust temperatures for galaxies at $z>5$ available in the literature compared with the results from our method. Here we briefly discuss the assumptions used in our derivation. 

For the galaxies MACS0416-Y1, B14-65666 and A1689-zD1 we refer to the dedicated works discussing the application of our method in comparison to multiple-band SED fitting\footnote{The minor differences in the quoted $T_{\rm d}$ values arise form the change in the adopted dust model, see also Section \ref{method}} (MACS0416-Y1, B14-65666: \citealt{TdCIImetodo}, A1689-zD1: \citealt{bakx:2021}). We highlight that A1689-zD1 is the only $z>5$ galaxy for which a band-9 continuum detection, short-wards of the peak of FIR emission, is available. It is very promising that also in this case -where  traditional SED fitting is particularly precise thanks to the widespread continuum data available- our method gives a consistent $T_{\rm d}$ value (within $\sim 0.5\ \sigma$). 

For the remaining sources detected in both \OIII $88\mathrm{\mu m}$ and \CII $158\ \mathrm{\mu m}$ (all but HZ4-10), \cite{2021Vallini} derived the value of $(\kappa_{\rm s},Z)$, albeit with large uncertainties (on average: $10 \simlt \kappa_{\rm s} \simlt 80$ and  $0.2 \simlt Z/Z_{\odot} \simlt 0.4$; see Table 1 in the paper). For each individual source we assume a random distribution for $(\kappa_{\rm s},Z)$ around the mean value, with the dispersion corresponding to the uncertainty. For all of these sources the ratio $y=r_{\rm CII}/r_{\rm star}$ is also measured: $y = 1.8 \pm 0.4$ (J1211-0118 and J0217-0208, \citealt{2020Harik}), $y=8 \pm 2$ (A2744-YD4, \citealt{laporte19}). Finally, for HZ4-10, due to the lack of observational constraints on $(Z,\kappa_{\rm s},y)$, we rely on the same assumptions used for both the ALPINE individual galaxies and the REBELS galaxies, which are extensively described and motivated in Section \ref{apllreb}.

In all cases with our method we derive $T_{\rm d}$ values consistent with literature estimates well within $\pm 30\%$. The more discrepant case is represented by J0217-0208 for which we predict a warmer temperature, albeit consistent within the large uncertainty given from SED fitting ($ \sim 52\%$). 

We note that for this galaxy, assuming the dust model adopted here, from traditional SED fitting (relying only on the two detections at $120\mathrm{\mu m},158\mathrm{\mu m}$) one would deduce a very large dust mass $\log (M_{\rm d}/M_{\odot}) = 9.28 \pm 0.17$. Given the stellar mass estimated for this galaxy ($M_{\star} \sim 3 \times 10^{10\ \mathrm{M_{\odot}}}$ from \citealt{2020Harik}), such dust muss would imply a very large dust yield of $y_{\rm d} = 3^{+2}_{-1}\ \mathrm{M_{\odot}}$ per SN, which is not compatible with SNe dust production constraints \citep{2016A&A...587A.157B,Matsuura2019,lesniewska2019dust,Slavin2020}. Further ALMA observations at shorter wavelengths will help us understand weather the dust temperature of this galaxy has been underestimated (and thus the dust mass overestimated), which would reduce the tension with the $T_{\rm d}$ value derived with our method. An other possibility is that the stellar mass of this source has been underestimated, this would relax the requirements set by the condition on $M_{\rm d} < M_{\rm d, max} \propto M_{star}$ (see Section \ref{method}), allowing for larger dust masses and lower temperatures in our derivation. If neither of these possibilities is verified, alternative dust production scenarios might have to be invoked for this source (see the discussion in Section \ref{MdREB}). 

\begin{table}
  \begin{center}
  \caption{Measured dust temperatures for galaxies at $z>5$ available in the literature compared with the results from our method. We note that for the literature data, we always show the result derived from traditional SED fitting in the most recent reference. The only case in which we provide two estimates is that of A2744-YD4, as the latest one is obtained from ad hoc simulations rather then direct measurements \citep[2, ][]{behrens18}. These temperatures are the ones shown in Fig. \ref{Tdz} as grey triangles (literature data) and stars (our derivations). \textbf{References}: 1 \citep{laporte19}, 2 \citep{behrens18}, 3 \citep{Bakx20}, 4 \citep{Tamura19}, 5 \citep{big3drag}, 6 \citep{bowler2018obscured}, 7 \citep{bakx:2021}, 8 \citep{knud17}, 9 \citep{2015Natur.519..327W}, 10 \citep{2020Harik}, 11 \citep{dusttemp2020}, 12 \citep{Pavesi_2016,2019Pav}, 13 \citep{capak15}.
     \label{tabhz}
    }
    \begin{tabular}{l|c|c|c|c|r}
        \hline\hline
        \multicolumn{1}{c}{\textit{Derived}}& \multicolumn{1}{c}{ID\#} & \multicolumn{2}{c}{\textit{Literature}}\\
        \hline
$T_{\rm d}\ [K]$ & & $T_{\rm d}\ [K]$ & $z$ & Ref.\\\hline
$<75$ & A2744-YD4 & $> 55$ & $8.38$ & \textit{1}\\[1mm]
$<107$ & A2744-YD4 & $91 \pm 23$ & $8.38$ & \textit{2}\\[1mm]
$<116$ & MACS0416-Y1 & $>80$ & $8.31$ &\textit{3, 4}\\[1mm]
$69^{+20}_{-15}$ & B14-65666 & $48-61$ & $7.15$ & \textit{5, 6}\\[1mm]
$40_{-7}^{+13}$ & A1689-zD1 & $47^{+15}_{-9}$ & $7.133$ & \textit{7, 8, 9}\\[1mm]
$38^{+11}_{-6}$ & J1211-0118 & $38^{+34}_{-12}$ & $6.0295$ & \textit{10}\\[1mm]
$49^{+11}_{-10}$ & J0217-0208 & $25^{+19}_{-5}$ & $6.204$ & \textit{10}\\[1mm]
$48^{+6}_{-8}$ & HZ10 & $46^{+16}_{-8}$ & $5.657$ & \textit{11, 12, 13}\\[1mm]
$49^{+17}_{-13}$ & HZ4 & $57^{+67}_{-17}$ & $5.544$ & \textit{11, 12, 13}\\[1mm]
$55^{+13}_{-14}$ & HZ9 & $49^{+29}_{-11}$ & $5.541$ & 
\textit{11, 12, 13}\\[1mm]
$37^{+14}_{-8}$ & HZ6 & $41^{+18}_{-7}$ & $5.293$ & \textit{11, 12, 13}\\[1mm]

      \hline\hline

    \end{tabular}
  \end{center}
\end{table}

% Don't change these lines
\bsp	% typesetting comment

\label{lastpage}
\end{document}